# Overview of the Massive Young Star-Forming Complex Study in Infrared and X-ray (MYStIX) Project

Eric D. Feigelson[*,1,2], Leisa K. Townsley[1], Patrick S. Broos [1], Heather A. Busk[1], Konstantin V. Getman[1], Robert King[3], Michael A. Kuhn[1], Tim Naylor[3], Matthew Povich[1,4], Adrian Baddeley[5,6], Matthew Bate[3], Remy Indebetouw[7], Kevin Luhman[1,2], Mark McCaughrean[8], Julian Pittard[9], Ralph Pudritz[10], Alison Sills[10], Yong Song[5], James Wadsley[10]

## ABSTRACT

MYStIX (Massive Young Star-Forming Complex Study in Infrared and X-ray) seeks to characterize 20 OB-dominated young clusters and their environs at distances $d \leq 4$ kpc using imaging detectors on the Chandra X-ray Observatory, Spitzer Space Telescope, and the United Kingdom InfraRed Telescope. The observational goals are to construct catalogs of star-forming complex stellar members with well-defined criteria, and maps of nebular gas (particularly of hot X-ray emitting plasma) and dust. A catalog of MYStIX Probable Complex Members (MPCMs) with several hundred OB stars and $> 30,000$ low mass pre-main sequence is assembled. This sample and related data products will be used to seek new empirical constraints on theoretical models of cluster formation and dynamics, mass segregation, OB star formation, star formation triggering on the periphery of H${\scriptstyle\rm II}$ regions, the survivability of protoplanetary disks in H${\scriptstyle\rm II}$ regions. This paper give an introduction and overview of the project, covering the data analysis methodology and application to two star forming regions, NGC 2264 and the Trifid Nebula.

[*]edf@astro.psu.edu

[1]Department of Astronomy & Astrophysics, Pennsylvania State University, 525 Davey Laboratory, University Park PA 16802

[2]Center for Exoplanets and Habitable Worlds, Penn State University, 525 Davey Laboratory, University Park PA 16802

[3]Department of Physics and Astronomy, University of Exeter, Stocker Road, Exeter, Devon, EX4 4SB, UK

[4]Department of Physics and Astronomy, California State Polytechnic University, 3801 West Temple Ave, Pomona, CA 91768

[5]Mathematics, Informatics and Statistics, CSIRO, Underwood Avenue, Floreat WA 6014, Australia

[6]School of Mathematics and Statistics, University of Western Australia, 35 Stirling Highway, Crawley WA 6009, Australia

[7]Department of Astronomy, University of Virginia, P.O. Box 400325, Charlottesville VA 22904

[8]Research & Scientific Support Department, European Space Agency, ESTEC, Postbus 299, 2200 AG Noordwijk, The Netherlands

[9]School of Physics and Astronomy, University of Leeds, Woodhouse Lane, Leeds, LS2 9JT, UK

[10]Department of Physics, McMaster University, 1280 Main Street West, Hamilton ON, L8S 4M1, Canada

Subject headings: infrared: stars; stars: early-type; open clusters and associations: general; planetary systems: protoplanetary disks; stars: formation; stars: pre-main sequence; X-rays: stars

# 1. Introduction

## 1.1. Star Formation in Giant Molecular Clouds

Recent decades have witnessed considerable progress in characterizing and understanding star formation that occurs in small molecular clouds. Millimeter and infrared studies of the Taurus, Perseus, Chamaeleon and similar nearby clouds give a detailed view of the phases of gravitational collapse, protostar formation, and early stellar evolution of stars in isolation or in small groups. The census of young stars in these regions is traced from intermediate-mass stars to cool brown dwarfs. Astrophysical modeling of isolated star formation is also in a well-developed state.

However, our understanding of star formation in massive star forming regions (MSFRs), particularly the emergence of rich star clusters, is more primitive with fundamental issues unresolved. Rich clusters are an important, perhaps the dominant, mode of star formation in the Galaxy (see reviews by Lada & Lada 2003; Allen et al. 2007; Kennicutt & Evans 2012). Questions of interest here include:

1. What are the essential conditions for rich cluster formation and how do these conditions arise in a molecular cloud complex? What are the effects of the new OB stars on cloud ionization and dispersal? (Magneto)hydrodynamical calculations of massive star formation in turbulent clouds are underway, but these are computationally expensive and are restricted to a narrow range of initial conditions (Peters et al. 2010; Moeckel & Bate 2010; Cunningham et al. 2011).

2. Do all stars in a cluster form essentially simultaneously during a single crossing time (Elmegreen 2000), or is star formation continuous active for millions of years (Tan et al. 2006)? Is it mainly a global process, or do rich clusters develop from the merger of smaller groups (McMillan et al. 2007; Maschberger et al. 2010)? Is the IMF constant during cluster formation both spatially and temporally (Krumholz et al. 2010)? Many clusters show a spread in apparent ages in the Hertzsprung-Russell diagram, but the interpretation of this effect is debated (Baraffe et al. 2009; Hosokawa et al. 2011; Jeffries et al. 2011).

3. Is the triggering of star formation in cloud material by expanding HII regions a major, or a minor, mode of star formation? Observational support for the 'radiatively driven implosion' model of triggered star formation near HII regions is growing (Ogura et al. 2007; Getman et al. 2009, 2012).

4. What are the principal mechanisms of massive OB star formation? Proposals include monolithic collapse, stellar mergers, rapid disk accretion, and 'competitative accretion' of ambient gas (Zinnecker & Yorke 2007). Some clear cases of disk accretion have been found (Cesaroni et al. 2007).

5. What are the causes of mass segregation, the concentration of OB stars in dense cluster cores? Does it arise from primordial star formation processes or rapid dynamical evolution (Allison et al. 2009)? Why is mass segregation absent in some young clusters (Wang et al. 2008)? Are many OB stars subject to dynamical ejection from cluster cores (Pflamm-Altenburg & Kroupa 2006)? Why are OB stars sometimes found to form after the lower mass stars (Feigelson & Townsley 2008; Ojha et al. 2010)?

6. Is the interstellar space within HII regions around rich clusters principally filled with photoionized gas at $10^4$ K temperatures, as assumed for a classical Strömgren sphere, or with hotter plasma at $10^7$ K from the shocked winds of OB stars? Several convincing cases of X-ray emitting plasma suffusing the interior of large HII regions have been found (Townsley et al. 2011b).

7. Are the conditions within rich star clusters hostile to the persistence of protoplanetary disks, suppressing planet formation? Photoionization and ablation of disks are seen in Orion Nebula proplyds (Johnstone et al. 1998).

## 1.2. Motivation for MYStIX

We believe that, in large part, these issues remain poorly understood because the endeavor is starved of high-quality data on stellar populations in MSFRs. While rich young clusters in the nearby Galaxy have been known for decades or centuries, the census of their stellar members has been very incomplete. Except for the nearest rich cluster, the Orion Nebula Cluster where the full Initial Mass Function (IMF) is clearly identified (except perhaps for ejected massive stars; Pflamm-Altenburg & Kroupa 2006), the census of stellar members for most young clusters is typically restricted to the dense central regions. In many cases, only a handful of lightly obscured OB stars are catalogued.

The weakness of the stellar membership census in and around massive clusters can be attributed to three difficulties encountered in optical and infrared surveys: spatially variable nebular line emission from heated gas and PAH emission from heated dust in the HII region and photodissociation region; spatially variable obscuration from the molecular cloud hosting the cluster; and crowding and contamination from Galactic field stars unrelated to the MSFR. For MSFRs in the Galactic Plane at low Galactic longitudes, there may be a hundred times more field stars than member stars in the magnitude range of interest. As a result, optical and infrared samples are largely restricted to MSFR members with distinctive photometric colors: lightly obscured OB stars with blue colors, and young stars hosting protoplanetary disks with infrared excesses. These restrictions miss the population of disk-free, lower mass young stars that often dominate the populations of young clusters.

These difficulties are significantly alleviated if sensitive, high-resolution X-ray images of the cluster and its environs are available. NASA's Chandra X-ray Observatory has proved to be a highly effective telescope for discriminating young stars in star forming regions up to distances of 4 kpc, even through obscuration of $A_V$ tens of magnitudes (Feigelson 2010). The emission processes are magnetic reconnection flares for lower mass pre-main sequence stars and shocked stellar winds for O stars. Contamination by older Galactic field stars is much reduced in X-ray images compared to optical or infrared images, and contamination by extragalactic sources can be mostly removed as the infrared counterparts are fainter than the pre-main sequence stars.

We describe here the Massive Young Star-Forming Complex Study in Infrared and X-ray (MYStIX) that combines the virtues of multiwavelength selection of young stellar populations − optical band for OB stars, infrared bands for young stars with protoplanetary disks, and the X-ray band for OB stars and pre-main sequence flaring stars. Uniform analysis procedures are applied wherever possible to the targeted MSFRs. Effort is exerted to obtain high-sensitivity catalogs from the X-ray and infrared images using advanced algorithms designed to treat the nebular and crowding problems. Probabilistic catalog matching and source classification algorithms give objective selection of cluster members; incompleteness and selection biases are still present but are reduced to acceptable levels for many purposes. Once cleaned of unresolved sources, the X-ray images then reveal the diffuse $10^7$ K plasma from shocked OB stellar winds or supernova remnants that fill some HII regions. Many of the analysis procedures are based on our earlier multiwavelength survey of the Carina Nebula complex (Townsley et al. 2011a).

The MYStIX project is principally based on analysis of archival X-ray data from NASA's Chandra X-ray Observatory, near-infrared (NIR) from the United Kingdom InfraRed Telescope (UKIRT) and the 2MASS survey, and mid-infrared (MIR) observations from NASA's Spitzer Space Telescope. Many of the UKIRT observations were obtained by the United Kingdom Infrared Deep Sky Survey Galactic Plane Survey (UKIDSS; Lawrence et al. 2007).

An important aspect of the MYStIX project is to compare properties of different star forming regions and young clusters. While past studies have investigated individual MSFRs, they are based on diverse methodologies and datasets that hinder intercomparisons. Although MYStIX clusters do not constitute a well-defined sample, they do represent a range of properties. Some are dominated by only one late-O star with M $\simeq$ 30 $M_\odot$ while others have dozens of O stars with masses up to M $\simeq$ 100 − 150 $M_\odot$. Some are embedded deep in cloud material while other lie at the edges of clouds or have entirely dissipated nearby cloud material. Some are relatively isolated structures on 1 − 5 pc scales while others are part of multi-cluster star formation complexes on 20 − 50 pc scales. Some have very high central star densities while others are more diffuse. Some have high protoplanetary disk fractions and others have low fractions or spatial gradients in disk fraction.

With 31,550 identified 'MYStIX Probable Complex Members' (MPCMs) in the 20 targeted MSFRs (Broos et al. 2013), the MYStIX project gives the largest samples of stellar populations of rich star forming regions obtained to date. Together with images of nebular emission, they provide a strong new empirical basis for addressing the astrophysical questions outlined in §1.1. This empirical foundation allows us to address questions concerning the structure and early evolution of clusters, the IMF and total stellar populations, total OB population (including obscured members), spatial gradients in stellar mass and protoplanetary disk distributions, protostellar populations embedded in cloud material around clusters, and relationships between hot, warm and cold gases in and around HII regions[1].

---

[1]In addition to their importance to Galactic star formation studies, our local Sun and Solar System likely formed in or near an OB-dominated cluster (Adams 2010; Dauphas & Chaussidon 2011). A recent analysis of the astrophysical origins of the short-lived radionuclides in Solar System meteorites suggests that the Sun formed with a few hundred stars in a molecular cloud several parsecs from an OB-dominated cluster with 1200 stars (Gounelle & Meynet 2012).

### 1.3. Outline of this paper

This paper provides an overview of the motivation, data sources, analysis methodology, and data products for the MYStIX project. It is published simultaneously with six papers that give detailed information and electronic data products for several steps in the project's progress. These papers present: single-wavelength analysis of Chandra X-ray data for 10 MYStIX fields (Kuhn et al. 2013a), UKIRT near-infrared (King et al. 2013), and Spitzer (Kuhn et al. 2013b) mid-infrared data; matching between X-ray and infrared sources (Naylor et al. 2013); identification of the 'MYStIX Infrared Excess Star' sample (MIRES; Povich et al. 2013); classification of X-ray sources and construction of the MYStIX Probable Complex Member sample (MPCM; Broos et al. 2013). Additional Chandra data are presented by Townsley et al. (2013). Several scientific studies will quickly emerge concerning: spatial clustering of MPCM stars (Kuhn et al., in preparation); a new stellar age estimator and star formation histories in MYStIX regions (Getman et al., in preparation); and candidate new OB stars in MYStIX regions (Busk et al., in preparation). A range of additional studies addressing the astrophysical issues outlined above are underway or planned.

## 2. The MYStIX Star-Forming Complex Sample

### 2.1. Sample selection

The Chandra X-ray Observatory mission has imaged about three dozen Galactic young rich stellar clusters and their vicinities. We have selected 20 of these with the following criteria: observed during the first decade of the mission, have nearby distances (d $\leq 4$ kpc), have young estimated ages (t$\leq 5$ Myr), and exhibit rich populations dominated by at least one O star. We further restrict the sample to those with Chandra exposures of sufficient duration to give X-ray limiting sensitivity $\log L_x \leq 30.0$ erg s$^{-1}$ in the hard $2-8$ keV band at the center of the Chandra field.

Note that we do not set a criterion based on obscuration or close association with a molecular cloud. Some of our targets are in the Lada & Lada (2003) catalog of nearby embedded clusters while others are in the Kharchenko et al. (2005) catalog of visible OB associations. An obscuration criterion is omitted because the X-ray emission of many (although not all) young stars is often hard enough to penetrate through considerable interstellar materials, $N_H$ $10^{22}$ cm$^{-2}$ hydrogen column density equivalent to tens of visual magnitudes assuming standard gas-to-dust ratio. NIR and MIR emission from protoplanetary disks can also be detected through very high column densities.

The MYStIX MSFRs are listed in Table 1 with a number of basic properties: common names, equatorial and Galactic coordinates, estimated or measured distance from the Sun, and earliest known spectral type. They are listed in order of increasing distance from the Sun. There is no consistent nomenclature for MSFRs; common names sometimes refer to the star cluster and sometimes to the associated gas as either a radio H$_{II}$ region or an H$\alpha$ emission nebula. We will use the names given in the first column of Table 1 in MYStIX studies, recognizing that more precise identifiers of cluster and cloud components will often be needed. The final column in Table 1 lists the major reviews on these clusters from the Handbook of Star Forming Regions (HSFR, Reipurth & Schneider 2008). Appendix A gives brief summaries of the MYStIX MSFRs, emphasizing studies published after these reviews were written.

The distance criterion d $\leq$ 4 kpc is needed to give sufficient sensitivity and resolution in the X-ray to resolve large numbers of individual stars in each cluster. This criterion eliminates the MSFRs associated with the starburst around the Galactic Center, and also removes the most luminous young 'super star clusters' in the Galaxy such as Westerlund 1 and NGC 3603 (see review by Turner 2009). However, by including mosaics of contiguous Chandra fields (each Chandra exposure subtends $17'\times 17'$ or $\sim$ 10 pc at d $\sim$ 2 kpc), the sample does include some star forming complexes tens of parsecs across comparable to 'giant extragalactic H$_{II}$ regions' seen in nearby spiral galaxies. These include the large survey of the Carina Nebula Complex described by Townsley et al. (2011a, and associated papers), portions of the W3/W4/W5 complex, and the less well-studied NGC 6334/NGC 6357 complex.

This MYStIX sample of massive star forming regions is not formally complete in any way. The sample is spatially restricted to only a few percent of the Galactic disk. But it is also restricted by the haphazard process of different scientists proposing, and different telescope allocation committees approving, targets to be observed with the Chandra X-ray Observatory during the first decade of the mission. This sample incompleteness restricts the nature of the scientific inferences we can make. Relationships between cluster properties (e.g., mass segregation vs. cluster central density, disk fraction vs. OB population) and classifications (e.g., clouds around clusters with and without triggered star formation, H$_{II}$ regions with and without hot plasma) can be studied. But the sample cannot be used to count how many Galactic clusters have high or low values of a given property or how many star forming regions fall into a given class. We can compare the stellar IMF within clusters, but we do not learn reliable information about the cluster mass function of the Galaxy.

## 2.2. Star Forming Region Distances and Ages

The distances to rich young clusters and their associated star forming regions have historically been difficult measure accurately. Recently, multi-epoch VLBI astrometric measurements of maser spots associated with high-mass protostars (or for the Orion Nebula region, VLBI measurements of low-mass non-thermal radio stars) give direct trigonometric parallax distances for some star forming regions. In the MYStIX sample, accurate VLBI parallactic distances are available for the Orion Nebula, DR 21, M 17, and W 3 (references in Table 1). We assume that the associated star forming complex has the same distance as the measured embedded star. We further assume that the Flame Nebula in the Orion B molecular cloud lies at the same distance as the Orion Nebula in the Orion A cloud, and that W 4 lies at the same distance as its adjacent W 3 cloud.

For the other clusters, distances are estimated by a variety of techniques: fitting of the high-mass main sequence on the Hertzsprung-Russell diagram (HRD), extinction maps of background stars, constraints from the X-ray luminosity function, or kinematic fitting of molecular cloud radial velocities. These estimates are often subject to considerable uncertainties. For example, fitting the ZAMS to OB stars may require determination of binarity and differential absorption for individual stars. Cluster stellar memberships are mainly main sequence OB stars where the locus in the HRD is nearly vertical so distance is poorly constrained. For low mass pre-main sequence stars, stellar age and distance are degenerate in the HRD. Systematic uncertainties can be larger than internal uncertainties for each distance estimation method. Tothill et al. (2008), for example, describe over a dozen estimates for the Lagoon Nebula with most in the range $1.3 - 1.8$ kpc. Cases where the distance appears to be obtained with greater accuracy (such as 1.48 kpc for NGC 2362) may simply represent fewer efforts to estimate the distance.

MYStIX science results are based on the distances listed in Table 1. Some findings will be vulnerable to distance errors. Comparisons of cluster sizes in units of pc, stellar surface densities in stars $pc^{-2}$, X-ray luminosities in erg $s^{-1}$, and IMFs could be systematically biased if the estimated distances are incorrect.

Table 1.  MYStIX Young Star-Forming Complexes

| Name | Alt name | Location | | Distance | | Earliest | SFR Handbook |
|------|----------|----------|---|----------|---|----------|--------------|
| | | $(\alpha, \delta)$ | $(l, b)$ | kpc | Ref. | Sp. Ty. | |
| (1) | (2) | (3) | (4) | (5) | (6) | (7) | (8) |
| Orion Nebula | M 42, NGC 1976 | 0535.3−05.4 | 209.0−19.4 | 0.414±0.007 | 1 | O7 | Muench et al. (2008) |
| Flame Nebula | NGC 2024, W 12 | 0541.7−01.9 | 206.5−16.4 | 0.414 | ⋯ | O8: | Meyer et al. (2008) |
| W 40 | RCW 174, Sh 64 | 1831.5−02.1 | 28.8+03.5 | 0.5 | 2 | late-O | Rodney & Reipurth (2008) |
| RCW 36 | VMR C | 0859.0−43.7 | 265.1+01.4 | 0.7±0.2 | 3 | O8 | Pettersson (2008) |
| NGC 2264 | Cone, Fox Fur | 0642.0+09.9 | 203.0+02.2 | 0.913±0.1 | 4 | O7 | Dahm (2008a) |
| Rosette Nebula | NGC 2244, W 16 | 0631.7+05.0 | 206.3−02.1 | 1.33±0.05 | 5 | O4 | Román-Zúñiga & Lada (2008) |
| Lagoon Nebula | M 8, NGC 6530 | 1803.6−24.4 | 6.0−01.2 | $1.3^{+0.5}$ | 6 | O7 | Tothill et al. (2008) |
| NGC 2362 | OCl 633 | 0718.7−24.9 | 238.2−05.6 | 1.48 | 7 | O9 I | Dahm (2008b) |
| DR 21 | W 75 | 2039.0+42.3 | 81.7+00.5 | 1.50±0.08 | 8 | ⋯ | Rygl et al. (2012) |
| RCW 38 | G268.0-01.0 | 0859.8−47.5 | 268.0−01.0 | 1.7±0.9 | 9 | O5 | Wolk et al. (2008) |
| NGC 6334 | Cat's Paw | 1720.0−36.0 | 351.1+00.7 | 1.7 | 10 | O8: | Persi & Tapia (2008) |
| NGC 6357 | Pis 24, W 22 | 1724.5−34.2 | 353.0+00.9 | 1.7 | 11 | O3.5 | Persi & Tapia (2008) |
| Eagle Nebula | M 16, NGC 6611 | 1818.8−13.8 | 17.0+00.8 | 1.75 | 12 | O9.5 | Oliveira (2008) |
| M 17 | NGC 6618, W 38 | 1820.8−16.2 | 15.1−00.7 | 2.0±0.1 | 13 | O4 | Chini & Hoffmeister (2008) |
| W 3 | IC 1795 | 0227.0+61.9 | 133.9+01.1 | 2.04±0.07 | 14 | O5 | Megeath et al. (2008) |
| W 4 | IC 1805 | 0232.7+61.5 | 134.7+00.9 | 2.04 | ⋯ | O4 | Megeath et al. (2008) |
| Carina Nebula | Tr 14/15/16 | 1044.3−59.9 | 287.6−00.9 | 2.3 | 15 | O2, LBV | Smith & Brooks (2008) |
| Trifid Nebula | M 20, NGC 6514 | 1802.7−23.0 | 7.0−00.3 | 2.7±0.5 | 16 | O7.5 | Rho et al. (2008) |
| NGC 3576 | RCW 57 | 1111.8−61.3 | 291.3−00.7 | 2.8 | 17 | O3-6? | ⋯ |
| NGC 1893 | IC 410 | 0522.8+33.4 | 173.6−01.7 | 3.6±0.2 | 18 | O5 | ⋯ |

Note. — Distance references: 1. Menten et al. (2007) 2. Shuping et al. (2012) 3. Baba et al. (2004) 4. Baxter et al. (2009) 5. Lombardi et al. (2011) 6. Tothill et al. (2008) 7. Moitinho et al. (2001) 8. Rygl et al. (2012) 9. Schneider et al. (2010) 10. Russeil et al. (2010) 11. Wang et al. (2012) 12. Guarcello et al. (2007) 13. Xu et al. (2011) 14. Hachisuka et al. (2006) 15. Smith (2006) 16. Cambrésy et al.

The situation is even more uncertain when the ages of MSFRs, or their constituent star clusters, are considered. For example, to estimate an age of 1.5 Myr for NGC 2264, Baxter et al. (2009) combine fitting pre-main sequence isochrones on the HRD with a revised distance. In W 4, Wolff et al. (2011) estimate an age of $1-3$ Myr from the HRD of massive stars. In cases where spectroscopy of individual stars is not available, cluster ages are sometimes estimated from isochrone fitting to color magnitude diagrams (e.g., Prisinzano et al. 2011, for NGC 1893). However, this method can systematically underestimate true cluster ages (Naylor 2009).

Nearly all HRD and color-magnitude diagrams of young clusters show an age spread, but it is difficult to interpret whether it represents a true range in birth times or a combination of other causes such as observational uncertainties, stellar variability, unresolved binaries, and different accretional histories (Baraffe et al. 2009; Reggiani et al. 2011; Jeffries et al. 2011; Littlefair et al. 2011). Even in the Orion Nebula Cluster, which has the best-characterized stellar population and distance of any rich young stellar cluster, this debate has not been resolved, and the problem is worse when comparing other clusters with uncertain distances. Finally, some MSFRs are studied nearly exclusively at long wavelengths for very young protostars, so that older populations can be missed. This problem can be alleviated by X-ray surveys because the magnetic flaring of pre-main sequence stars is elevated for hundreds of millions of years. For example, an X-ray study revealed populous older clusters lying in front of the NGC 6334 cloud complex that is primarily known for its massive protostars (Feigelson et al. 2009), and Chandra has revealed a significant older, widely distributed older stellar population in the Carina Complex (Townsley et al. 2011a).

Given these difficulties, until individual stellar components of the MSFRs are studied in detail, we do not assign specific ages to each MYStIX star forming region. Many have active star formation today associated with ages $< 0.1$ Myr; some have massive supergiants or a truncated upper main sequence indicating ages $\geq 5-10$ Myr. Typical ages for most cluster members are usually around $1-3$ Myr. Three of the MYStIX clusters – NGC 2362, Tr 15 in Carina, and the northern cluster of NGC 3576 – are arguably older with ages $\geq 5$ Myr, as they have no associated molecular material and are missing the more massive O stars, likely due to supernova explosions (Moitinho et al. 2001; Wang et al. 2011).

Table 2.   MYSTIX Datasets

| Name | Chandra | | Near-IR | Mid-IR | |
|------|---------|---|---------|--------|---|
| | Pointings | Publs. | | SST | Publs. |
| (1) | (2) | (3) | (4) | (5) | (6) |
| Orion[a] | 1 (838) | ⋯ [a] | ⋯ [a] | ⋯ [a] | 38, 39 |
| Flame | 1 (80) | 1, 2 | 2MASS | archive | 38 |
| W 40[b] | 1 (40) | 3 | ⋯ [b] | archive | 40, 41 |
| RCW 36 | 1 (75) | ⋯ | 2MASS | archive | |
| NGC 2264 | 3 (100+50+60) | 4, 5, 6, 7 | UKIRT | archive | 42, 43, 44, 45, 46 |
| Rosette | 5 (100+4x20) | 8, 9, 10, 11 | UKIDSS | archive | 47, 48 |
| Lagoon | 2 (61+180) | 12, 13 | UKIDSS | GLIMPSE | 49 |
| NGC 2362 | 1 (100) | 14, 15, 16 | UKIRT | archive | 50, 51 |
| DR 21 | 1 (100) | ⋯ | UKIDSS | archive | 52, 53, 54 |
| RCW 38 | 1 (100) | 17, 18, 19 | 2MASS | Vela-Carina | 19 |
| NGC 6334 | 2 (40+40) | 20, 21 | UKIRT | GLIMPSE | 55 |
| NGC 6357 | 3 (40+40+40) | 22 | UKIRT | GLiMPSE | 56 |
| Eagle | 3 (80+80+80) | 23, 24, 25, 26 | UKIDSS | GLIMPSE | 25, 57, 58 |
| M 17 | 2 (340+100) | 8, 27, 28 | UKIDSS | GLIMPSE | 59, 60 |
| W 3 | 4 (3x80+50) | 29, 30, 31 | 2MASS | archive | 31, 61, 62 |
| W 4 | 1 (80) | ⋯ | 2MASS | archive | 31 |
| Carina[c] | ⋯ [c] | 32, 33, 34 | ⋯ [c] | Vela-Carina | 63, 64, 65, 66, 67 |
| Trifid | 1 (60) | 35 | UKIDSS | GLIMPSE | 68, 69, 70 |
| NGC 3576 | 2 (60+60) | 29 | 2MASS | archive | ⋯ |
| NGC 1893 | 1 (446) | 36, 37 | UKIRT | archive | 33 |

[a]Chandra Orion Ultradeep Project. X-ray source list and properties were obtained from Getman et al. (2005). About 22 publications based on the X-ray sources are associated with Getman et al. (2005) in the October 2005 Special Issue of the Astrophysical Journal Supplements. The near-infrared counterparts are based on deep VLT/ISAAC observations. Mid-infrared counterparts are obtained from the Spitzer survey of Megeath et al. (2012).

## 3.   MYStIX datasets

### 3.1.   Chandra X-ray Data

Table 2 lists the multiwavelength datasets used for the MYStIX project. The first columns summarize the archived Chandra observations examined. ObsIDs (column 2) gives the observation identifiers. 'Pointings' (column 3) indicates the number of different exposures. Each exposure subtends $17' \times 17$ using Chandra's four-CCD Advanced CCD Imaging Spectrometer Imaging Array (ACIS-I, Garmire et al. 2003). The following numbers in parenthesis gives the approximate exposure time in kiloseconds. For typical Chandra exposures of star forming regions around  2 kpc, and applying the well-established (though poorly understood) correlation between X-ray luminosity and stellar age for pre-main sequence stars (Preibisch et al. 2005; Telleschi et al. 2007), Chandra's sensitivity is sufficient to capture most 1 $M_0$ stars and higher-mass stars with a decreasing fraction of lower mass stars. We thus emphasize that, although Chandra images typical show hundreds to thousands of pre-main sequence members, these represent only a small portion of the full IMF that peaks around 0.3 $M_0$.

For seventeen MYStIX targets, we reanalyzed archived Chandra data using the procedures outlined in §5.1. Methodology, X-ray source lists and images for these regions are presented by Kuhn et al. (2013a) and Townsley et al. (2013). For three MYStIX MSFRs − Orion Nebula Cluster, W 40, and the Carina Complex − X-ray analysis at the same level of sensitivity had been carried out by our group (Getman et al. 2005; Kuhn et al. 2010; Broos et al. 2011a). In these cases, we used the published X-ray source lists and stellar properties for MYStIX analysis.

Column (4) lists the principal published studies of the Chandra image of MYStIX star forming regions. Most involve the stellar populations, but a few discuss diffuse plasma X-ray emission. Most of the X-ray images have been studied in the past, some in considerable detail. We reiterate that the role of the MYStIX project is to analyze the full sample with the same methodology available today − the most sensitive point source detection techniques and careful characterization of diffuse emission. The existing publications are heterogeneous, and often not optimal, in X-ray analysis capabilities.

### 3.2.   UKIRT and 2MASS Near-Infrared Data

A critical component of the MYStIX project is the availability of JH K imaging of most targeted MSFRs with the Wide Field Camera on the United Kingdom InfraRed Telescope (UKIRT). Many of these observations were performed by the United Kingdom Infrared Deep Sky Survey Galactic Plane Survey (UKIDSS GPS) surveying 1800 sq.deg. of the Galactic Plane (Lawrence et al. 2007; Lucas et al. 2008). The remaining fields were observed with the same camera using UKIDSS procedures. In regions where crowding is unimportant (such as NGC 2264), the UKIRT sensitivity reaches J'''19.6, H'''18.9 and K'''17.9 at a signal-to-noise of 10. Typical UKIDSS point spread functions (PSFs) have $0.8^{\parallel} - 1.0^{\parallel}$ FWHM with $0.4^{\parallel}$ pixels. For finding counterparts of Chandra sources in MYStIX fields, spatial resolution is often more important than photometric depth, both to resolve crowded stellar fields and to reduce the effect of spatially variable nebular emission in HII regions associated with the clusters.

For MYStIX targets inaccessible to UKIRT, and for brighter stars that are saturated in UKIRT images, we obtained JH K photometry from the Point Source Catalog of the Two Micron All Sky Survey (2MASS PSC) obtained with dedicated 1.3m telescopes (Skrutskie 2006). 2MASS sensitivity limits are approximately J ''' 15.8, H ''' 15.1, and $K_s$ ''' 14.3. For the more distant MYStIX targets, the 2MASS PSC is inadequate for identifying young stellar counterparts to Chandra sources both due to limited sensitivity and spatial resolution. With $2^{\parallel}$ pixels, 2MASS source blending difficulties start around $6^{\parallel} - 8^{\parallel}$ separations (Cutri 2006, §4b). However, for the nearer MYStIX star forming regions (hence with brighter and more widely separated pre-main sequence members), 2MASS has fewer difficulties.

The near-infrared coverage of the Chandra MYStIX targets is listed in column (5) in Table 2. This can be summarized as follows: eleven star forming regions have corresponding UKIDSS or UKIRT observations, three have published near-infrared counterparts to high-quality X-ray source lists, and six have 2MASS survey coverage. The UKIRT observations are presented in detail by King et al. (2013).

### 3.3. Mid-Infrared Data

The Infrared Array Camera (IRAC) on the Spitzer Space Telescope provides wide-field imaging in four MIR bands: 3.6μm, 4.5μm, 5.6μm and 8.0μ (Fazio et al. 2004). At the shorter wavelengths, blackbody photospheres of even fainter cluster members are readily seen at the distances of MYStIX star forming complexes. At the longer wavelengths, dust emission from protoplanetary disks dominate. Used by itself, IRAC data provide samples of disk-dominated young stars if unrelated Galactic field stars can be removed. Used in conjunction with X-ray selection, IRAC complements UKIRT in providing stellar counterparts to X-ray emitting young stars. The MYStIX approach analyses spectral energy distributions (SEDs) with the seven bands combined from UKIRT and IRAC to identify young disk-bearing stars (Povich

et al. 2013).

All of the MYStIX star forming complex targets have Spitzer IRAC coverage (last column of Table 2). Five are only covered by the short-exposure Galactic Legacy Infrared Mid-Plane Survey Extraordinaire (GLIMPSE) project (Benjamin et al. 2003). Ten have deeper archived observations that we analyze as part of the MYStIX effort. For the Carina complex, we take infrared results published by the Chandra Carina Complex Project (Townsley et al. 2011a; Povich et al. 2011b; Preibisch et al. 2011). For RCW 38 and NGC 3576, the data are obtained from the Vela-Carina survey (Majewski, PI).

## 4. Two Prototype MYStIX Star-Forming Regions: NGC 2264 and the Trifid Nebula

To illustrate the methodology and scientific potential of the MYStIX project, we focus on two MYStIX star forming complexes that exemplify range of challenges encountered in this effort. The presentation here concentrates on the identification of stellar members of the MSFRs, a sample we call 'MYStIX Probable Complex Members' (MPCMs). The MYStIX project also involves mapping of diffuse X-ray emission, and comparing it to maps of cloud gas and dust. This aspect is not discussed here for the prototype MSFRs.

NGC 2264 is relatively nearby with bright cluster members and it lies in a region with low field star contamination. Its pre-main sequence stellar membership has been studied for decades in the visible band. However, the stars are not concentrated into a single rich cluster, and the region is dominated by late-O and early B stars that produce relatively weak nebular HII regions. The Trifid Nebula is more distant and lies in a very crowded region of the Galactic Plane at longitude $7°$. It has a centrally concentrated main cluster dominated by an O6 star that ionizes a very bright HII region nebular emission. Its pre-main sequence membership is poorly established with no study in the visible band. Both NGC 2264 and Trifid have both revealed young stars and active star formation in molecular clouds; a wide range of absorptions is thus present. The NGC 2264 and Trifid regions are briefly reviewed in §A.5 and §A.18, respectively. This section is limited to a qualitative examination of multiwavelength images of the prototype clusters before quantitative MYStIX analysis has taken place. The following section (§5) traces the MYStIX data and science analysis procedures to show the results of the MYStIX analysis for these prototype MYStIX MSFRs (§6).

## 4.1. NGC 2264

Figure 1 shows the NGC 2264 region at X-ray (Chandra), near-infrared (UKIDSS), mid-infrared (Spitzer) and visible (Digital Sky Survey) wavelengths. (Note that we consider visible light images in this section although they are not included in MYStIX analysis.) The polygon outlines the Chandra mosaic that constitutes the field of MYStIX study. The images have qualitatively different characteristics. The visible image shows stars, but the population is severely limited by obscuring dust clouds and, to a lesser degree, by Hα nebulosity. The near-infrared image penetrates deeper and gives a more reliable stellar census; both obscuration and nebulosity (mostly Brγ emission) are reduced. The mid-infrared image shows many of these stars plus others in heavily obscured cloud cores but it is more severely affected by bright nebulosity, mostly from emission bands of polycyclic aromatic hydrocarbons, and by saturation of the brightest infrared protostars. In all of the optical and infrared images, the large majority of stars are Galactic field stars distributed throughout the image, giving a challenge to selection of recently formed stars. The X-ray image has no immediately evident emission nebulosity (the diffuse orange structures Figure 1 are due to the detector background), and the sources are a combination of cluster members and background extragalactic sources. Comparison with the infrared images shows that X-ray detections are common even among cluster members with moderately heavy absorption. With many fewer contaminants, the X-ray image list gives a more direct indication of the young stellar population and structure than the infrared images.

A closer examination of some distinctive locations gives further insight into the contributions and limitations of each waveband.

1. In the northern ACIS field, the V = 4.6 O7 main sequence star S Mon lies on the edge of a small ($\sim 30''$ $\sim$ 0.1 pc) clump of pre-main sequence stars seen in X-rays and infrared. These lower mass members are invisible in standard visible light images due to the wings of the O star PSF.

2. The bright 'Fox Fur Nebula' LBN 912 $\sim 10'$ southwest of S Mon has an amorphous morphology in visible Hα images, but appears as a prominent incomplete shell-like H II region in the Spitzer image. Associated with a small molecular cloud (Teixeira et al. 2012), it is likely illuminated by the off-center B1 star BD +09°1331B, an isolated massive star without an associated concentration of X-ray emitting pre-main sequence stars.

3. The embedded massive star IRS 2 lies near the top of the southern ACIS field lies in curved and elongated stellar structure $\sim 2'$ long with several dozen members seen in both infrared and X-rays. Each band has advantages and deficiencies. The mid-infrared shows very heavily absorbed faint stars but is insensitive in the inner $\sim 30''$ due to bright dust emission around IRS 2. The near-infrared image shows the central object is a visual binary with separation $\sim 2''$ and sees some of the associated young stars. The X-rays from the massive member is relatively weak (as is typical for intermediate-mass A and B stars) but X-ray selection here does not add significantly to the youngest embedded populations.

4. The IRS 1 region above the Cone Nebula is the most crowded portion of the X-ray image. As with the IRS 2 region, the cluster has an irregular structure without a central concentration, extending $\sim 10'$ to the south. IRS 1 itself is not detected in X-rays, but a number of closeby X-ray stars are masked by diffuse emission and saturation in the infrared images.

5. The $V = 7.2$ B2 III star HD 47887, famous as the bright star $2'$ north of the bright tip of the Cone Nebula, is a weak X-ray source with several X-ray pre-main sequence in its immediate vicinity. The two optical stars superposed on the ionized portion of Cone Nebula are brightly detected in X-rays.

## 4.2. Trifid Nebula

Figure 2 shows a four-band view of the Trifid Nebula MYStIX field, delineated by a single Chandra ACIS pointing (upper left panel). Here the OB stars produce a high-surface brightness HII region nebular emission that prevents detection of the lower mass population in the visible band (lower right panel). The near-infrared image has much reduced nebulosity, but the dense stellar pattern is nearly uniform across the field (lower left panel). This indicates that the vast majority of UKIDSS stars are Galactic field stars unrelated to the star forming regions because the Trifid, lying in a nearby Galactic spiral arm, is projected against the Galactic Center region. The mid-infrared image (upper right panel) shows nebulosity from heated dust (mostly PAH bands), and a moderately rich stellar field that is also nearly uniform. Thus, superficial examination of all long-wavelength images does not reveal the young stellar cluster.

1. Near the top of the Chandra field of view, the Spitzer image shows a north-facing bright rimmed cloud. Nearly 50 X-ray luminous stars lie above this cloud; more would likely be resolved if the region were on-axis. The brightest is the intermediate-mass $(K = 8.0)$ Class II young star 2MASS J18025044-2247501 (Rho et al. 2008). This loose collection of young stars has not previously been recognized as a young stellar subcluster.

2. The northern component of the Trifid Nebula, characterized by a large reflection nebula of blue light in the visible image (Figure 2, lower right), is illuminated by the A7 I supergiant HD 164514. It is isolated without an associated cluster of infrared or X-ray emitting lower mass stars.

3. The main (southern) ionized component of the Nebula is ionized by the double O6 star HD 164492 lying near the intersection of the three dark dust lanes. The X-ray image reveals a moderately rich cluster distributed asymmetrically $5'$ east and north of the dominant star (Rho et al. 2004).

4. Infrared-excess protostars are concentrated in dense cloud cores, particularly to the south of the emission nebula (Rho et al. 2008).

## 5.  MYStIX Data Analysis Methodology

Figure 3 shows a diagram of the principal data analysis tasks involved in MYStIX analysis. We briefly describe these methods in the following subsections. Details and electronic source lists are given in accompanying MYStIX papers on the Chandra X-ray observations (Kuhn et al. 2013a; Townsley et al. 2013), UKIRT near-infrared observations (King et al. 2013), Spitzer mid-infrared observations (Kuhn et al. 2013b), X-ray/infrared source matching (Naylor et al. 2013), infrared excess determination (Povich et al. 2013), and MSFR membership classification (Broos et al. 2013). We illustrate the application of these methods on the prototype clusters, NGC 2264 and Trifid.

### 5.1.  X-ray analysis

MYStIX analysis of the archived Chandra data is based on the ACIS Extract (AE) package and associated recipes developed by the ACIS Instrument Team at Penn State since 2002. Written in the Interactive Data Language, AE is described in detail by Broos et al. (2010). The procedure has a variety of advantages over standard Chandra data analysis tools. Source candidates are found a local bumps in a maximum likelihood reconstruction of the image using local PSFs; this method is more sensitive and reliable than commonly used procedures based on the wavelet transform (Townsley et al. 2006). The final source list is constructed iteratively in the original image of photon events based on a statistical significance level that a source exists above a local Poisson background. The use of a local (rather than global) detection criterion allows consistent detection when the background rate varies due to overlapping exposures and/or diffuse astrophysical emission. Source photons are then extracted using accurate models of the telescope PSF. The X-ray photometry of faint sources are estimated nonparametrically using the source counts and median energy following procedures described by Getman et al. (2010). Diffuse X-ray emission is analyzed by extracting and smoothing source-free regions of the image, and then proceeding with parametric spectral analysis following procedures described by Townsley et al. (2011b). The MYStIX analysis is similar the analysis of the Chandra Carina Complex Project (CCCP) with a few enhancements. Figure 4 shows unsmoothed X-ray images and source extraction regions for portions of the NGC 2264 and Trifid fields.

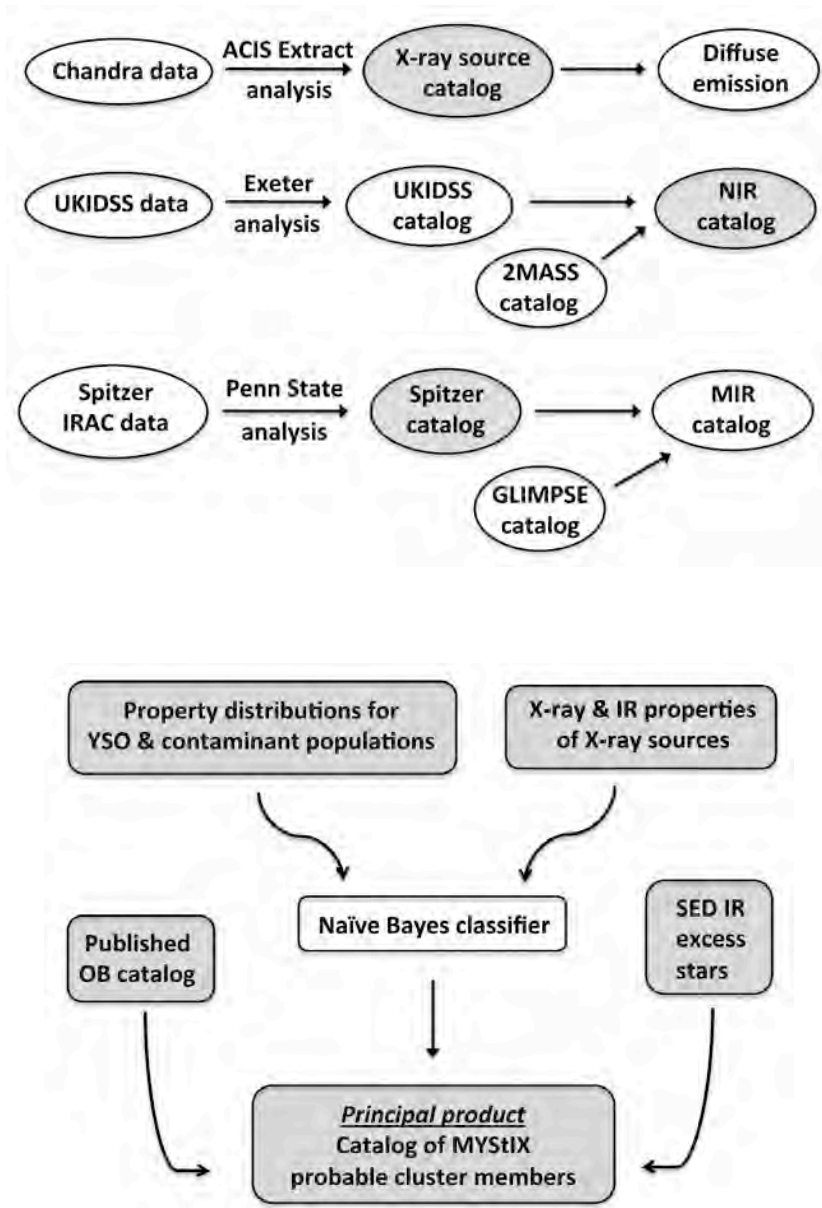

Fig. 3.— Flow diagram for analysis of MYStIX clusters. The top portion outlines the analysis of the data in each waveband: X-ray, near-infrared, and mid-infrared. The bottom portion outlines the combination of multiwavelength sources to produce a probabilistic list of cluster members. Shaded boxes represent electronic source tables provided in accompanying MYStIX papers.

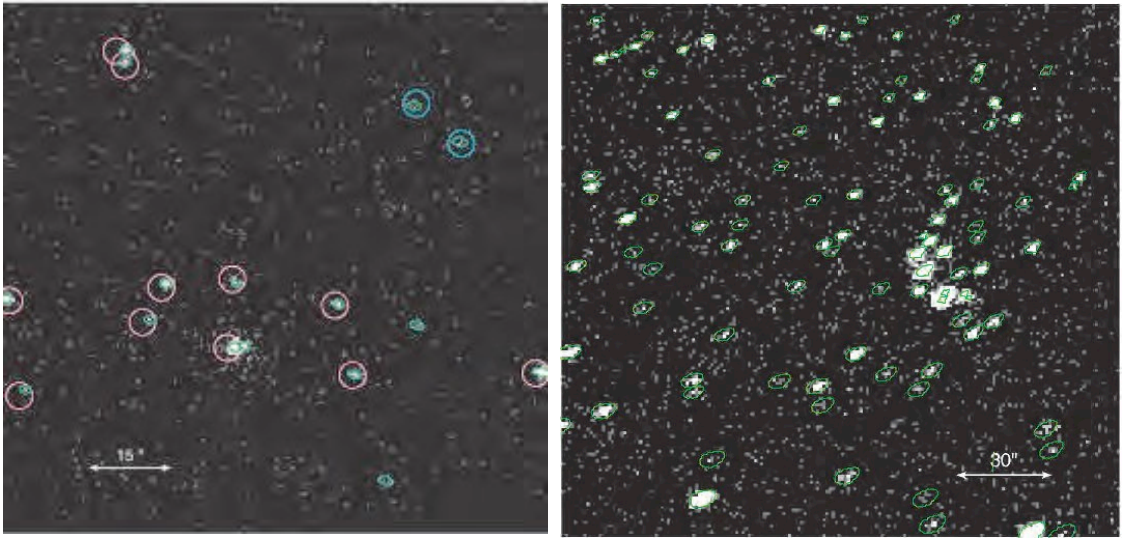

Fig. 4.— Example of Chandra images illustrating challenges in MYStIX source detection. Left: Northern portion of the NGC 2264 ACIS mosaic. Green polygons show the ACIS Extract source extraction regions, magenta circles are sources found by Ramírez et al. (2004), and cyan circles are two additional sources found by Sung et al. (2004). Right: Dense source concentration in the Trifid Nebula.

The MYStIX source detection procedures are considerably more sensitive than some other commonly used analysis procedures for ACIS data, partly due to lower source existence thresholds and partly due to improved treatment of local background levels and crowded source regions. This increase in sensitivity is important for MYStIX star formation region memberships because the X-ray luminosity function of pre-main sequence stars rises rapidly from the high luminosity limit around $\log L_x \approx 32$ erg s$^{-1}$ to 30.0 ergs s$^{-1}$, the typical sensitivity limit of a MYStIX exposure (see Figure 9 of Wang et al. 2008). Thus an improvement of a factor of 2 in faint source sensitivity can give a factor of $2-3$ more cluster member detections. This was a central motivation for the reanalysis of archival Chandra cluster data.

The sensitivity of our source detection method is illustrated in Figure 4, (left panel) where MYStIX finds two faint isolated sources ( 5 photons) and resolve two close doubles (separation $1.0 - 1.5''$) missed by earlier researchers. In this ACIS exposure, the Chandra Source Catalog lists 239 sources, Ramírez et al. (2004) found 263 sources, and Sung et al. (2004) found 271. The MYStIX analysis, in contrast, locates 450 sources in this pointing. One can also see that the PSF centroiding method in ACIS Extract sometimes gives more accurate source positions than wavelet positions.

The MYStIX X-ray analysis procedures, and resulting X-ray source lists, are given by Kuhn et al. (2013a) and Townsley et al. (2013).

## 5.2. Near-Infrared Analysis

The UKIRT Wide Field Camera (WFCAM, Casali et al. 2007) performs wide-field surveys in the JH K bands. MYStIX uses data from both the UKIDSS Galactic Plane Survey (Lawrence et al. 2007) and independent WFCAM observations. MYStIX analysis is based on procedures developed for UKIDSS (Hodgkin et al. 2009) modified for the crowding and nebulosity typically found in MSFRs. For example, source extraction is made with smaller apertures. Photometry for bright saturated stars is replaced with 2MASS photometry. The combination of UKIRT catalog with 2MASS photometry for bright stars results in the MYStIX NIR catalog.

Figure 5 (left panel) shows a small portion of the UKIDSS K-band image around the bright stars ionizing the Trifid Nebula. Visual examination of the MYStIX star catalog shows it is highly sensitive and reliable for finding well-separated stellar sources. But in high star density (low Galactic longitude) and high nebulosity regions, a number of errors are seen. False positives can appear as spurious sources in saturated PSFs of bright stars, near intermediate brightness stars (yellow arrow in Figure 5), in regions of bright emission nebulosity, and along sharp gradients in nebular emission (cyan arrows). False negatives include some missed stars  $1''$ from stars (red arrows and cyan X), missed faint stars in dense obscuring clouds with rapid surface brightness gradients, and missed stars in bright filamentary nebulosity. The orange arrows show detector effects which do not trigger source detection by our algorithm. In the most crowded fields, these problems are a few percent of the UKIDSS sample. The catalog has not been cleaned or altered to treat these problems except in the vicinity of X-ray sources. Solin et al. (2012) give a detailed description of the challenges and successes of locating young stellar clusters from UKIDSS Galactic Plane

False sources are seen in the PSFs of bright sources (red circles with black arrows) and occasionally in diffuse patches (cyan arrow). The algorithm misses some faint sources appearing mostly in the 3.6 μm band (green arrows). We provide a second catalog based on more sensitive source detection settings that captures many of these faint sources; however, these settings also give rise to increased false positive sources.

Overall, the MIR source detection algorithm appears to be effective with good sensitivity and few false positives even in regions of spatially variable nebular emission. Of course, the sensitivity to faint stars is unavoidably greatly reduced in the close vicinity of the brightest infrared sources such as IRS 1 in NGC 2264. The MYStIX procedures are comparable in sensitivity to the GLIMPSE analysis procedure with very similar photometry. Comparing to the MYStIX mid-infrared catalog to that obtained by Sung et al. (2009) for the same NGC 2264 dataset shows our catalog includes 96% of their sources and has 21% additional faint sources.

A full description of the mid-infrared analysis and source lists for the MYStIX project is given in the accompanying paper by Kuhn et al. (2013b).

## 5.4. X-ray/Infrared Source Matching

The matching of X-ray to infrared source catalogs is particularly tricky for the MYStIX project. First, the positional uncertainties of Chandra sources are 'heteroscedastic', depending on location in the field and count rate. Uncertainties can range from $0.2''$ on-axis to $5''$ far off-axis. The infrared positional uncertainties are constant across the field, though they can increase for fainter sources. Second, Spitzer images have lower resolution than UKIDSS or Chandra on-axis images, so unique one-to-one correspondences with mid-infrared sources may not be possible in crowded regions.

Third, and most important, is the frequent dominance of Galactic field stars over MSFR members. For uncrowded MYStIX complexes like NGC 2264, the Galactic field star contamination is not heavy and a straightforward positional matching based on these positional uncertainties can give reliable X-ray/infrared counterparts. But for distant clusters, particularly those projected near the Galactic Center like the Trifid, $90-99\%$ of stars may be field stars. A single Chandra or Spitzer source can have multiple UKIDSS counterparts so that false matches to foreground or background stars become common. This effect is illustrated in Figure 6 (left panel).

We thus develop a statistical matching algorithm that accounts for the heteroscedastic measurement errors and introduces a weighting that favors matches to UKIRT stars having magnitudes characteristics of the young stellar population under consideration in each field. This method is based on the work of Sutherland & Saunders (1992). It is used in MYStIX for both Chandra-UKIRT and Chandra-Spitzer matching. Figure 6 (right panel) shows the result of its application to the crowded Trifid Nebula region; if the infrared source closest to an X-ray source is faint, it may be rejected as the true counterpart in favor of a more distant, but brighter, source. Naylor et al. (2013) fully describe the method and provide lists of infrared matches to MYStIX X-ray sources.

## 5.5. Infrared Excess Sources

Populations of young stars in clusters and star forming regions have long been obtained by detecting photometric infrared excesses from dusty protoplanetary disks orbiting pre-main sequence stars. Young stellar objects have traditionally been identified either in polygonal regions of infrared color-color diagrams (e.g., Grasdalen et al. 1973; Lada & Adams 1992; Allen et al. 2004) or by fitting star-plus-disk models to infrared spectral energy distribution (e.g., Robitaille et al. 2007). Unfortunately, for the more challenging MYStIX clusters that have bright, spatially variable nebulosity and many thousands of Galactic field stars unrelated to the star forming region, straightforward application of established infrared excess criteria give samples that are clearly dominated by non-cluster members. Problems arise from contributions of dusty galaxies seen through the Galactic Plane, dusty post-main sequence asymptotic giant branch (AGB) stars, faulty photometry from nebular knots incorrectly identified as stars, and from incorrect matching of near- and mid-infrared stars. As a result, the MYStIX procedures involve a number of additional criteria to reduce contaminants at the expense of completeness.

Sources in the UKIRT (§5.2) and Spitzer(§5.3) catalogs are first merged using a simple proximity rule (not the the magnitude-weighted matching outlined in §5.4). Seven photometric values constitute the spectral energy distribution (SED) are typically available for each infrared sources in the MYStIX fields (J, H, K, 3.6 μm, 4.5 μm, 5.6 μm, and 8.0 μm bands). A range of star-plus-disk models are fit to these SEDs using by weighted least squares regression following the procedures of Robitaille et al. (2007) after systematic photometric errors are added to the measurement errors. Spectral energy distributions for three of the 282 X-ray selected IR excess in NGC 2264 are shown in Figure 7 to illustrate the results of this analysis. Stars with a number of characteristics are omitted to reduce contaminants. These include sources in the IRAC color-magnitude diagram consistent with dusty galaxies; inadequate signal-to-noise in a sufficient number of spectral bands; and unphysical structure in the SED (usually associated with PAH nebular contamination). We designate the resulting sample culled of these problematic SEDs as 'MYStIX InfraRed Excess Sources' (MIRES).

The MIRES sample is still often dominated by unclustered sources unrelated to the star forming region. The surface density of remaining infrared-excess sources away from the known clusters is measured and assumed to represent remaining contamination by AGB stars and galaxies. A probability of cluster membership is then calculated for each source based on the local surface density of infrared-excess objects, and only sources above some probability threshold are flagged as probable cluster members (§5.7). The infrared-excess sample that enters the MPCM catalog is thus biased towards clustered groups of infrared-excess sources, and is less sensitive to widely dispersed populations.

A full description of the infrared excess analysis with tabulated MIRES source lists is given in the accompanying paper by Povich et al. (2013). For many MYStIX fields, this analysis is performed over a wider field of view than covered by the Chandra exposures. The MIRES sample thus has infrared excess stars that are not included in the MPCM sample.

## 5.6. Published OB Stars

The catalog of cluster members includes all stars in the MYStIX fields of view that are identified as OB stars by optical spectroscopy. We obtain these stars by combining stars listed in two collations of the astronomical literature: the catalog of stellar spectral classifications by Skiff (2010) and the SIMBAD astronomical database[2]. Stars with spectral types B3 through O2 are included. More modern types are used when discrepant classifications are present in the historical literature.

As historical positions of OB stars often do not have the subarcsecond accuracy needed for MYStIX analysis, we use positions from our near-infrared catalog. Historical positions are matched to $JHK$ band catalogs, and a prominent near-infrared star is typically found within $1''$. As these stars are bright in the near-infrared, positions and magnitudes are typically obtained from the 2MASS catalog rather than from UKIRT data where the image and photometry can be badly saturated. But the low resolution of the 2MASS telescope can be confused by the crowded environment and/or binary components of high-mass systems.

The association of X-ray sources to published OB stars presents several particular problems. It is not uncommon for two or more X-ray stars to lie within the 2MASS point spread function of massive stars (see for example, the O5 star HD 46150 in Rosette; Wang et al.

---

[2]http://simbad.u-strasbg.fr/simbad

2008). To mitigate these problems, X-ray sources close to published OB stars were examined in the Chandra and UKIRT images, and the literature of the OB stars was studied for reliability of the spectral classification. A number of associations that were not obtained by the automated system were added. Results from these examinations of OB stars are given in the footnotes of the MPCM lists given by Broos et al. (2013).

## 5.7. Statistical Classification of 'MYStIX Probable Complex Members'

Although the X-ray and infrared-excess source populations of MYStIX fields show clear concentrations of stellar clusters, it is not trivial to reliably associate an individual source with the young stellar population. In simpler single-waveband situations, straightforward decision trees can be effectively used, such as 'Disk-bearing young stars can be discriminated from disk-free stars by Spitzer IRAC colors [3.6]-[4.5$\geq$0.0 and [5.8]-[8.0]$\geq$0.4' or 'An X-ray source exhibiting a strong variations on timescales of hours is a magnetically active young star'. Our multiwavelength study not only needs to combine criteria such as these, but it needs to reduce large contaminating populations of Galactic field stars.

To address this challenge in classifying MYStIX sources, we build upon the statistical classifier for X-ray sources developed by Broos et al. (2011b) for the Chandra Carina Complex Project (CCCP). Based on 'naive Bayes classifiers', the probability that an X-ray source lies in a chosen class is treated as the product of independent probabilities associated with different properties (J band magnitude, mid-infrared colors, X-ray hardness and variability, and so forth). The method requires prior knowledge of the properties from 'training sets', giving in advance the probability distribution of each property for each class of target and contaminants. The construction of training sets for both young stars and contaminant populations for the CCCP is presented by Getman et al. (2011) and further refined in Broos et al. (2013). The probability that an X-ray source is classified as a young star is increased when it lies in a localized spatial concentration of X-ray sources. This is based on the premise that contaminant populations (both stellar and extragalactic) will be roughly spatially uniform. Table 3 lists the four classes and eight properties used in the MYStIX classifier.

Five data products flow into the classifier (see also the bottom panel of Figure 3): the X-ray source catalog (§5.1); the catalog of infrared-excess stars, including both X-ray selected stars and the full infrared catalogs (§5.5); the catalog of published O and early-B stars in the MYStIX field (§5.6); spatial maps of expected young star and contaminant distributions; and the class likelihood functions of the properties in Table 3 for each class based on training sets or simulations. Figure 8 illustrates this with distributions J magnitudes of X-ray source counterparts, one of the principal discriminators in the classifier.

Table 3.  MYSTIX Source Classification

| | Classes of objects |
|---|---|
| H1 | Foreground Galactic stars |
| H2 | MYStIX cluster members |
| H3 | Background Galactic stars |
| H4 | Galaxies and active galactic nuclei |
| | Source properties |
| X-ray sources | J magnitude |
| | X-ray median energy[a] |
| | X-ray variability[b] |
| | X-ray source density map[c] |
| Infrared sources | [4.5] magnitude[d] |
| | SED infrared excess[e] |
| | MIR source density map[c] |
| Optical | Spectroscopic OB star |

[a]This measures line-of-sight absorption.

[b]This indicates magnetic flaring.

[c]This indicates spatial clustering.

[d]This discriminates extragalactic sources that are always faint.

[e]This indicates warm circumstellar dust.

After the 'naive Bayes' product of class probabilities is computed for each X-ray source, a decision rule is applied to define when the probability of MSFR membership exceeds the probability of a contaminant population. We also indicate when an X-ray source has so little associated information that no classification is feasible. Many of the fainter members of rich clusters are seen only at one or another band, and thus do not have multiwavelength confirmation of membership. Most contaminants are 'unclassified'.

X-ray sources satisfying these classification criteria are combined with infrared sources with SEDs satisfying the criteria for dusty disks (§5.5) and spectroscopic OB stars (§5.6) to constitute the final list of MPCMs.

The full description of the MYStIX source classifier and other elements entering the MPCM sample construction is given in the accompanying paper by Broos et al. (2013). The resulting MPCM samples seem effective in most respects, giving large populations of highly clustered stars, often with X-ray emitting stars dominating rich clusters and infrared-excess stars distributed in the molecular cloud around the main clusters. Spatially uniform star distributions that may be highly contaminated with non-MSFR populations have low surface densities.

## 6.    Results of MYStIX Analysis of the Prototype Star Forming Complexes

Table 4 and Figure 9 summarize the results at several stages of the MYStIX analysis for the prototype NGC 2264 and Trifid Nebula targets. Tabulations similar to Table 4 for the full MYStIX MSFR sample are given by Broos et al. (2013); an abbreviated version is shown in Table 5 here. The Chandra populations are generally in the range $500 - 3000$ sources for each MYStIX target. While we see here that the NGC 2264 field has roughly twice the X-ray population of the Trifid Nebula (line 1 of Table 4), this is largely a function of the closer distance, deeper exposures, and multiple pointings of NGC 2264 (Table 2) rather than an assertion that NGC 2264 intrinsically has twice the number of MPCM members as the Trifid. Similarly, the finding that NGC 2264 has seven published OB stars compared to only two in Trifid (line 4) similarly may not reflect the underlying populations; it is particularly difficult to locate blue stars in the Trifid region where the nebular emission and obscuration are strong and spatially complex. The MYStIX project will provide new lists of candidate OB stars from its MPCM catalogs (Busk et al., in preparation). Careful evaluation of sensitivity limits (which vary across the field due both to intra-pointing degradation of the PSF and to inter-pointing exposure differences) is needed before total stellar populations can be estimated from MPCM samples.

The NIR source population in the Chandra field of view (line 2) is 7 times higher, and the MIR population (line 3) is 2.5 times higher, in the Trifid Nebula compared to NGC 2264 despite the smaller field of view of the Trifid. This illustrates the enormous contamination by Galactic field stars in MYStIX fields at low Galactic longitudes: sometimes >99% of the infrared stars have no relation to the star formation region under study. This problem has inhibited infrared-only studies of stellar populations in many rich star formation regions. The majority of X-ray sources ( 60%) have stellar matches in either or both of the NIR and MIR catalogs (lines $5-7$ of Table 4). The top panels of Figure 9 give more details on the X-ray matching results. In NGC 2264, the UKIRT and Spitzer surveys provide nearly identical matches: 95% of the near-infrared matches have mid-infrared counterparts, and vice versa. The availability of infrared photometry for most X-ray sources allows SED analysis showing that only $10-20\%$ of X-ray sources have infrared excesses (line 8 of Table 4). This confirms the long-standing experience (see review by Feigelson 2010) that Chandra is most effective at locating Class III disk-free pre-main sequence stars. It is clear that young star samples based only on infrared excesses often miss the majority of the young stellar populations in these fields.

Table 4.  Source Populations in Prototype MYSTIX Fields[a]

| Line | Population | NGC 2264 | Trifid |
|---|---|---|---|
| | **Single-wavelength results** | | |
| 1 | Chandra X-ray sources ' | 1,328 | 633 |
| 2 | UKIDSS/2MASS NIR sources | 11,865 | 76,251 |
| 3 | Spitzer MIR sources | 10,284 | 26,020 |
| 4 | Published OB stars | 7 | 2 |
| | **Multi-wavelength results** | | |
| 5 | X-ray/NIR matches [b] | 753 | 355 |
| 6 | X-ray/MIR matches [b] | 769 | 240 |
| 7 | X-ray/(NIR or MIR) matches [b] | 799 | 364 |
| 8 | MIRES[c] | 556 | 174 |
| | X-ray detected | 282 | 60 |
| | X-ray undetected | 274 | 114 |
| | **Classification results** | | |
| 9 | X-ray foreground stars[d] | 0 | 3 |
| 10 | X-ray background stars[d] | 0 | 10 |
| 11 | X-ray extragalactic objects[d] | 126 | 38 |
| 12 | X-ray young stars[d] | 898 | 418 |
| 13 | X-ray unclassified[d] | 304 | 164 |
| 14 | MPCMs[e] | 1173 | 532 |

[a]Spatially restricted to X-ray field of view

[b]Counterpart probability >0.80 using the magnitude-weighted proximity procedure (Naylor et al. 2013)

[c]MIRES = MYStIX InfraRed Excess Sources (Povich et al. 2013)

[d]Includes X-ray sources only

[e]MPCM = MYStIX Probable Complex Member, including classified X-ray sources, infrared SED excess sources, and spec-

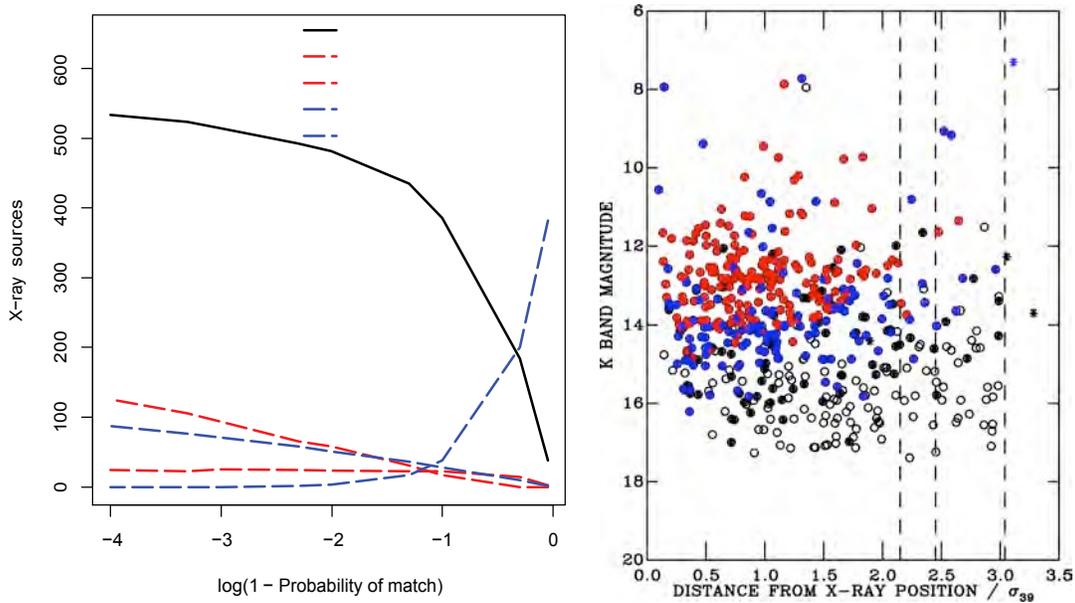

Fig. 6.— Results from two methods for matching Chandra X-ray sources with UKIDSS stars in the crowded Trifid Nebula field. Left: Proximity-only matching (Broos et al. 2010). The abscissa gives the radius allowed for matches, viewed as a probability associated with the X-ray error circle; smaller radii are towards the right. The ordinate gives the number of X-ray sources matched. The black curve shows the increase in matches as the radius allowed for matches increases towards the left. The associated changes of several types of matching errors are shown. Right: K-band magnitude weighted matching (Naylor et al. 2013). The abscissa is the offset distance between the X-ray and infrared sources scaled to the X-ray error circle radius. Open circles represents the nearest UKIDSS counterpart to each X-ray source. Red, blue and black circles are cases where the probability of being a true counterpart is >99%, 90%-99%, and 67%-90%, respectively, based on a weighting using the counterpart's K magnitude. Unfilled circles are rejected as counterparts in favor of brighter counterparts with wider separation from the X-ray position. The vertical dashed lines represent the 90%, 95% and 99% confidence regions based on the X-ray positions without consideration of K magnitudes.

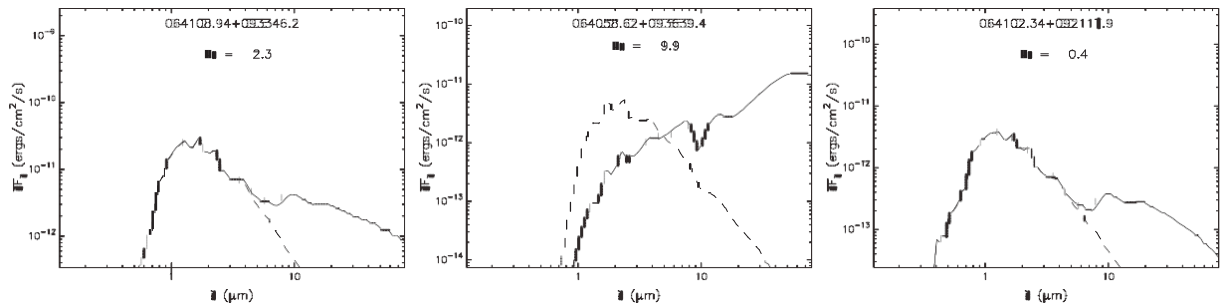

Fig. 7.— Three examples of X-ray sources with infrared excess from SED modeling in the NGC 2264 region: (a) previously studied T Tauri star; (b) previously identified Class I protostar; and (c) previously uncatalogued star with a weak infrared excess. The solid curves show the best-fit SED model and the dashed curves show the dereddened photospheric contribution from the best-fit star-plus-disk model.

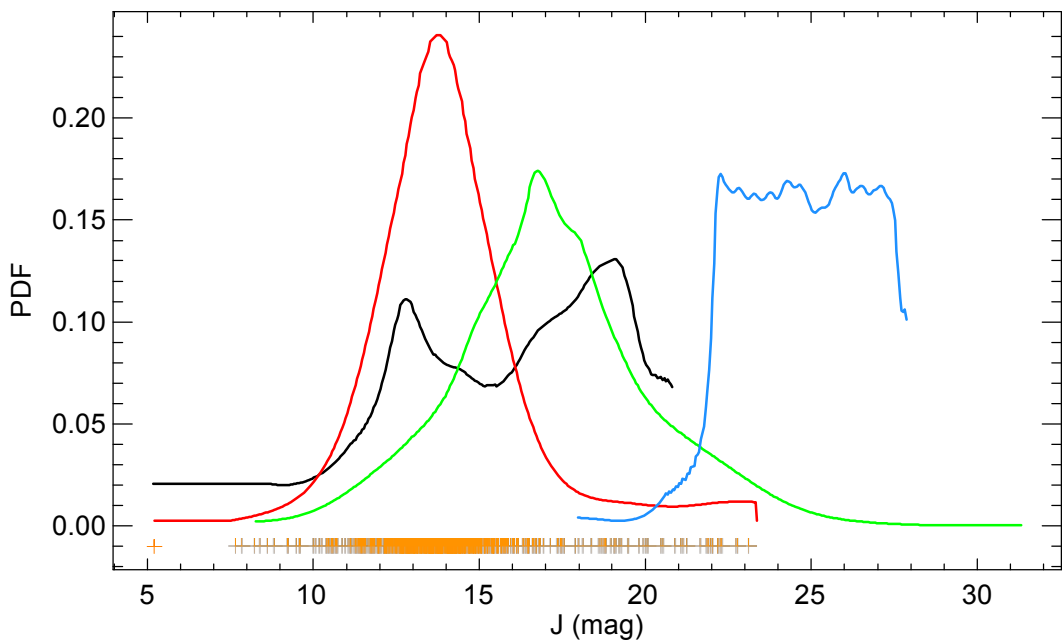

Fig. 8.— Distributions of the J-band magnitudes of the 'training set' samples in the NGC 2264 field illustrating the discriminating power for X-ray source classification. The red curve shows the distributions of observed X-ray sources in high-density (clustered) regions with individual sources are shown by orange plusses. The other curves are derived from simulations of contaminating populations (Getman et al. 2011; Broos et al. 2013): foreground Galactic stars (black), background Galactic stars (green), and extragalactic active galactic nuclei (blue). Each curve is normalized to encompass unity area.

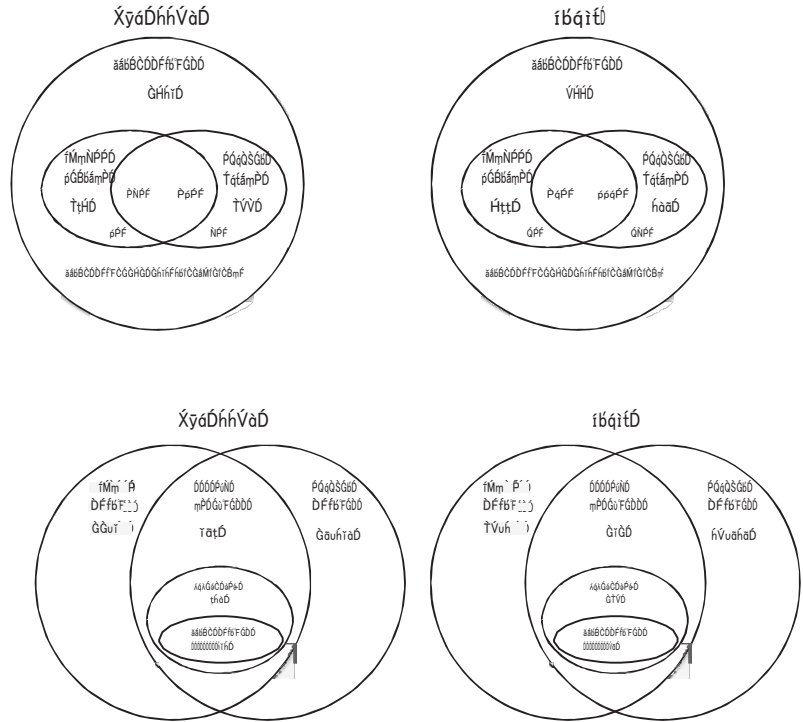

Fig. 9.— Venn diagrams showing intermediate results from MYStIX processing of the prototype NGC 2264 and Trifid Nebula fields. Top: X-ray sources with and without matched counterparts in the near-infrared UKIDSS and mid-infrared Spitzer images. Bottom: Infrared sources with and without photometric excesses at longer wavelengths, restricted to the X-ray field of view. The category 'SED IR excess' includes contaminants such as dusty galaxies, AGB stars, and spurious detections of nebular knots.

troscopic OB stars.

The results of the multi-property statistical classification of X-ray sources are summarized in lines $9-14$ of Table 4. Only a handful of sources are confidently classified as foreground or background Galactic field stars (lines $9-11$); simulations of the Galactic stellar X-ray populations indicate that most contaminants are among the X-ray sources with uncertain classifications (line 13). The young stars are indicates in lines 12 and 14: in NGC 2264 [Trifid Nebula], we find 898 [418] of the 1,328 [633] X-ray sources are probable members of the star forming region, and 1,173 [532] stars are probable members when infrared-excess and spectroscopic OB stars are added. Although the classifier combines all information in a complicated probabilistic fashion for the X-ray sources, one can roughly see that 75% of the members would have been identified by virtue of their X-ray emission alone, and 40% of the members would have been identified by virtue of their photometric infrared excess emission alone (compare lines 8, 12 and 14 of the table and the diagrams in Figure 9).

Further insight into the relationship between infrared and X-ray selection can be inferred from Figure 10 showing the MPCM population on spatial maps of NGC 2264 and the Trifid Nebula. In the northeast region of the NGC 2264 field, we see that the subcluster of X-ray selected stars around S Mon (item 1 in §4.1) are older without infrared disks (small yellow circles) while a clump of mostly disk-bearing stars (large red circles) lies off-center of the Fur Fox nebula to the west. In the southern portion of the field, rich clusters of very young members extend around and between IRS 1 and IRS 2. Here infrared-excess selection captures most of the members, but X-ray selection improves the sample where the infrared image suffers crowding or saturation. Finally, we do not see an obvious spatial gradient in MPCM surface density or disk fraction from inside to outside the Cone Nebula at the south of the region. This suggests that the Cone Nebula is not undergoing a burst of triggered star formation as seen in some other bright rimmed clouds (e.g. Getman et al. 2009).

The Trifid Nebula results (Figure 10b) also shows informative spatial segregation of disk-bearing and disk-free stars. A diffuse group of mixed disk-bearing and disk-free stars lie in the bright rimmed cloud to the north of the bright H$_{II}$ region (item 1 in §4.2). The rich cluster inside the nebula has a high fraction of older, disk-free stars. The densest concentration of members in the immediate vicinity of the heavily absorbed O star CD $-23°13804B$ have no disky members at all; disk destruction by the radiation and wind of the massive star is a possibility. Finally, a considerable population of disk-bearing stars is seen to the southwest of the bright nebula and extends beyond the X-ray field of view. These very young stars appear to be aligned with filamentary Infrared Dark Clouds (IRDCs). Absorptions from the handful of X-ray sources within or behind the IRDCs may give valuable measures of their gas columns.

Table 5. Summary Source Populations in All MYSTIX Fields[a]

| MYStIX | # Chandra | # MI RES | # MP CM |
|---|---|---|---|
| Orion | 1616 | 631 | 1524 |
| Flame | 547 | 193 | 485 |
| W 40 | 225 | 308 | 426 |
| RCW 36 | 502 | 135 | 384 |
| NGC 2264 | 1328 | 556 | 1173 |
| Rosette | 1962 | 622 | 1730 |
| Lagoon | 2427 | 468 | 2056 |
| NGC 2362 | 690 | 67 | 491 |
| DR 21 | 765 | 507 | 980 |
| RCW 38 | 1019 | 112 | 886 |
| NGC 6334 | 1510 | 407 | 1667 |
| NGC 6357 | 2360 | 523 | 2235 |
| Eagle | 2830 | 721 | 2574 |
| M 17 | 2999 | 155 | 2364 |
| W 3 | 2094 | 259 | 1676 |
| W 4 | 647 | 415 | 519 |
| Carina | 7412 | 815 | 7334 |
| Trifid | 633 | 174 | 532 |
| NGC 3576 | 1522 | 142 | 1213 |
| NGC 1893 | 1442 | 538 | 1301 |

## 7. Comparison with Independent Membership Surveys of NGC 2264

Among all MYStIX targeted MSFRs, NGC 2264 has the most comprehensive published catalogs of complex members[3]. We compare the MYStIX selection of 'probable cluster members' in NGC 2264 with two previous catalogs that are independent from the MYStIX datasets: the photometric variables of Walker (1956) which were historically important in establishing the pre-main sequence nature of T Tauri variables; and the sensitive Hα star survey of Dahm & Simon (2005). The individual stellar associations between MPCM stars and stars from these and other published membership surveys appear in the footnotes of the electronic MPCM membership table in the accompanying paper by Broos et al. (2013).

One hundred seventy three of the photometric variables studied by Walker (1956) lie in the X-ray exposures defining the MYStIX field of view for NGC 2264. Most (76%) of these are astrometrically matched with X-ray sources associated with near-infrared stars. The remainder are dominated by A and late-B stars which are often undetected in the X-ray band. The photometrically variable sample size is 16% of the MPCM sample of MPCM stars indicating that, at least with the precision achieved by Walker using photographic plates, photometric variability alone provides a very incomplete cluster sample.

Approximately 430 Hα stars of Dahm & Simon (2005, designated 'IfAHα' stars) lie in the MYStIX field of view of which 83% are recovered as MPCM stars. Dozens of other MPCM stars recover members found in the ESO-Hα survey (Reipurth et al. 2004). The failure to recover 17% IfAHα stars is largely attributed to the limited sensitivity of the X-ray and mid-infrared surveys; the missing low mass members are on average 1 mag fainter than the recovered stars. This loss is expected: due to the strong correlation between X-ray luminosity and stellar mass among pre-main sequence stars (Telleschi et al. 2007), X-ray surveys will not detect lower mass young stars in most MYStIX regions.

The MPCM sample is less effective in recovering the youngest members of NGC 2264. Forbrich et al. (2010) identify several dozen infrared-excess stars in the Spokes Cluster (IRS 2 vicinity) from Spitzer spectroscopy. While most (16 of 20) stars spectroscopically classified as Class II are in the MPCM catalog, only 5 of 14 Class I stars are found. The poor performance of MPCM in recovering Class I protostars is partly due to their heavy absorption of X-ray emission in the Chandra band: three of the five X-ray detected protostars have extremely high Median Energy $\simeq 4.5-5.8$ keV indicating $A_V > 100$ mag (see Fig 4 of Getman et al. 2010). Other X-ray sources with comparable absorption may have been erroneously place in the 'Unclassified' category due to the paucity of similar objects in the young star 'training set' (Broos et al. 2013). The conservative photometric selection criteria used to identify young stellar infrared excess sources may also have excluded some spectroscopic Class I stars (Povich et al. 2013). But the spectroscopic sample of Forbrich et al. is also very incomplete: our MIRES [MPCM] samples have 5 [ 8] times more stars in the small region around IRS 2 than they consider in their Spitzer spectroscopic study.

---

[3]The Orion Nebula Cluster also has an excellent membership list from prior near-infrared study, but its MYStIX dataset is anomalous: the Chandra observations are several times longer than for other targets (from the Chandra Orion Ultradeep Project, Getman et al. 2005) and the Spitzer observations are unusually insensitive due to the high surface brightness of the Orion Nebula throughout the field of view.

The MYStIX X-ray/infrared matching procedure performed very well: four cases (1%) of an X-ray source proximate to an Hα star failed to match to the UKIRT near-infrared star due to binarity or far-off-axis X-ray positional error, but all of these cases were recovered in matches to Spitzer UKIRT sources. The MYStIX statistical classification procedure was also successful; only one Hα star that was astrometrically matched to an X-ray source was not classified as a MPCM. The MPCM census has 340 X-ray sources matched to K band stars that are not detected in Hα, indicating that the Hα sample captures somewhat more than half of the lightly obscured young stars obtained by MYStIX. The missed members are mostly non-accreting Class III systems selected by their X-ray emission.

Of the 1,174 MPCM sources in the NGC 2264 field, 205 have no published association within $2''$ in the SIMBAD database. These can be considered completely new probable members of the NGC 2264 complex. Many lie in the eastern portion of the field that has been less well-studied in earlier surveys.

We thus find that MYStIX recovers about 80% of traditional optical-band variable star and Hα samples, and about 35% of a Class I protostar sample, using X-ray and infrared selection criteria. Unfortunately, Hα star surveys are not feasible for many MYStIX MSFRs where spatially varying HII region nebular Hα and cloud obscuration are prevalent. However, sensitive multi-epoch variability surveys of star forming regions such as the VISTA Via Lactea survey (§8) may give variable stars cluster member subsamples that can be added to MYStIX samples in the future.

## 8. Discussion: Laying the Empirical Foundation for Stellar Population Studies in Massive Star Formation Regions

The MYStIX approach to stellar populations of star formation complexes is essentially to join together X-ray selected stars with infrared-excess stars and spectroscopic OB stars into a single sample of "MYStIX Probable Complex Members" (MPCMs). A considerable number of other studies have taken a similar approach for individual clusters[4]. Our group has developed sophisticated methodology to address several tricky aspects of the effort, and applies these methods to 20 MSFRs to facilitate comparative studies to further our understanding of clustered star formation. The observational foundation of the MYStIX project described here is more fully presented in the accompanying MYStIX papers (Kuhn et al. 2013a; Townsley et al. 2013; King et al. 2013; Kuhn et al. 2013b; Naylor et al. 2013; Povich et al. 2013; Broos et al. 2013).

The analysis effort to construct MPCM samples can be viewed as a sequence of three stages (Figure 3). First, we collect single-wavelength images of each region from the archives of the Chandra X-ray Observatory, UKIRT and its UKIDSS project supplemented by the 2MASS surveys, and the Spitzer Space Telescope. We emphasize consistent analysis with methods carefully designed to identify sources in crowded and nebulous regions. Our X-ray analysis techniques, in particular, typically doubles the number of X-ray sources obtained by traditional procedures.

Second, we match the single-wavelength source catalogs with each other, taking into account source variations in positional measurement error and the likelihood that the match corresponds to a young stellar member. This reduces contamination by uninteresting field stars that can overwhelm proximity-only source matching procedure. The NIR-MIR matches are then subject to photometric SED analysis to find the infrared-excess stars likely associated with protoplanetary disks. Several decision rules, and a weight in favor of high local surface densities, are applied to reduce likely contaminants. The young stars entering the MPCM catalog from the MIRES infrared-excess catalog is thus based on very restrictive rules, and many infrared-excess stars are probably excluded from MPCM.

The third stage is a probabilistic classification of X-ray sources to identify likely MSFR members. This is needed to combine information from a variety of analysis efforts in an objective fashion. Some sources − such as published OB stars and X-ray sources with infrared excesses or flares − are very likely to be classified as MSFR members. In other cases, the chances of MSFR membership is based on the combined properties of the source compared to training sets of both members and contaminants, again with weighting to favor sources in spatially clustered regions. For many sources, the multiwavelength properties are too sparse for classification, and most of the contaminants (foreground or background

---

[4]A list of such studies is given in the review by Feigelson (2010). More recent studies include W 40 (Kuhn et al. 2010), Cyg OB2 (Wright et al. 2010), Eagle Nebula (Guarcello et al. 2010), IC 1795 (Roccatagliata et al. 2011), RCW 38 (Winston et al. 2011), NGC 1893 (Prisinzano et al. 2011), Cep OB3b (Allen et al. 2012), and the large Chandra Carina Complex Project by Townsley et al. (2011a). Some of these studies use near-infrared or mid-infrared, but not both bands, together with X-ray selection.

Galactic field stars, external galaxies or active galactic nuclei) are placed into an 'Unclassified' category.

Appendix B discusses the limitations of the resulting MPCM samples. In some ways, the samples are too small (false negatives), missing true young stars associated with the targeted MSFRs. In other ways, the samples are too large (false positives), adding to the MPCM samples that are not real members. It is difficult to evaluate the completeness of MPCM samples although progress is possible with subsamples, in particular by matching the observed X-ray luminosity function of X-ray selected members to the Orion Nebula Cluster (e.g., Getman et al. 2006). We note that similar difficulties in completeness evaluation are present in many traditional methods for identifying young stellar populations – such as optical variability, Hα emission, or infrared excess. Comparison with different selection techniques in the NGC 2264 region (§6-7) shows that the MYStIX multiwavelength samples are much more complete than samples based on infrared-excess or any other single property alone.

The MYStIX project also takes a particular approach to statistical decision making known as 'soft classification'. In matching X-ray sources to crowded infrared star fields, we calculate a probability of matching to plausible possible counterparts and a probability that no match is present (Naylor et al. 2013). An arbitrary decision rule based on these probabilities is then applied to make the X-ray/infrared counterpart assignment. In classifying X-ray sources as young stars or one of three populations of contaminants, we calculate a probability for each class based on several source properties and apply an arbitrary decision rule to make the class assignment (Broos et al. 2013). In both X-ray source classification and in infrared excess classification (Povich et al. 2013), we weight classification by the local spatial density of likely complex members. The final assignment decision rule (equivalent to setting a '3 sigma' detection criterion for faint source detection) can be easily revised by other researchers because we provide intermediate tabular results of class probabilities. Soft classifiers differ from commonly used hard classifiers that establish sharp classification boundaries and bypass class probability estimation. For example, Class 0-I-II-III classification of infrared excess using polygons in an infrared color-color plot is a hard classification technique. These results can not be revised later without new computations. Both soft and hard approaches are commonly used in modern statistical applications (Wahba 2002; Liu et al. 2011).

While the MYStIX project is a significant observational effort, it is incremental in various respects. Future deep exposures of MYStIX targets with the Chandra X-ray Observatory can add new X-ray sources to the MYStIX sample. Most of the existing Chandra exposures are too short to capture the bulk of the young stellar population; new exposures in the range 0.3− 1 Ms could increase the X-ray selected cluster membership several-fold without encountering confusion. In the near-infrared, several new capabilities have recently emerged. The 4m-class Visible and Infrared Survey Telescope for Astronomy (VISTA) telescope produces surveys in six bands over a large field of view with high resolution ($0.3''$ pixels; Emerson & Sutherland 2010). VISTA is now engaged in the Via Lactea project involving multiple scans of Galactic Plane fields to locate and characterize large populations of variable stars, including pre-main sequence stars (Minniti et al. 2010; Saito et al. 2012). The NOAO Extremely Wide Field Infrared Imager (NEWFIRM) camera for 4m-class telescopes covers near-infrared bands ($0.4''$ pixels; Probst et al. 2008). The High-Acuity Wide-field K-band Imager (HAWK-I) camera for ESO's Very Large Telescope has a smaller field of view but higher resolution ($0.1''$ pixels) and sensitivity than other available imagers (Kissler-Patig et al. 2008). We hope that, as higher quality X-ray and infrared observations become available for MYStIX clusters, our electronic tables of single-wavelength sources will be useful for counterpart searches.

The restriction of the MYStIX sample to MSFRs within distances $\leq 4$ kpc also eliminates the most massive and luminous 'super-star clusters' in the Galaxy including those in the nuclear starburst around the Galactic Center (see review by Turner 2009). In these star forming regions with bad crowding and high absorption, it is difficult to detect individual pre-main sequence stars and study has been mostly restricted to O stars and supergiants. Further investigation of these most massive star clusters in the Galactic Plane is possible with long Chandra exposures and high-resolution infrared imagery.

Finally, we note that the 'MYStIX Probable Complex Member' data product is based only on spatial, X-ray and infrared photometric properties that are combined into a probabilistic classification of membership. Spectroscopy, preferable in the near-infrared bands, is needed to solidify individual memberships. Multiobject $JHK$ spectrographs on southern sky telescopes are particularly needed for this task.

## 9. Conclusions

No single astronomical method can identify the full stellar population emerging over time from a massive star forming complex. The historically important tools of photometric variability and H$\alpha$ emission from accretion in the visible band capture different portions of the classical T Tauri population, and the selection hot blue stellar photospheres captures OB stars with low obscuration. However, these visible band survey techniques are not available for most massive star formation regions that are often subject to both spatially varying cloud obscuration and nebular emission contamination. Infrared excess and submillimeter surveys locate the youngest stars with dusty protoplanetary disks. However, Galactic field star contamination in optical and infrared images often overwhelm efforts to identify the full young star population. X-ray surveys reveal the disk-free pre-main sequence stars down to a stellar mass correlated with the X-ray sensitivity limit, as well as OB stars and significant portions of the disk-bearing population. Here field star contamination is much reduced. Each waveband captures a portion of the full stellar population, often overlapping with samples obtained at other wavebands (Figure 9).

The best available approach to the stellar census of star formation regions is to combine methods from different wavebands. The MYStIX project combines the capabilities of three instruments − Chandra X-ray Observatory ACIS spectroscopic imager, the UKIRT wide-field near-infrared imager in three bands, and the Spitzer Space Telescope IRAC mid-infrared imager in four-bands − with historical OB stars to obtain new lists of cluster members in 20 OB-dominated star forming complexes at distances between 0.5 and 4 kpc. The full list of 31,550 MPCM stars in 20 MYStIX MSFRs is given in Broos et al. (2013).

Specialized source detection techniques are used to achieve high sensitivity (particularly in the X-ray images where sources with as few as 3 photons are identified) while treating crowding and nebular contamination. Our source detection philosophy is to produce the most sensitive single-wavelength source lists, even at the expense of false positive detections, and then cull the lists by applying quantitative criteria to multiwavelength properties. As infrared images are often overwhelmed by older Galactic field stars, the selection of cluster members is greatly boosted by the detection of X-ray emission and/or infrared excess from a protoplanetary disk. Statistical methods are applied to multiwavelength source matching in crowded fields, and to discriminate cluster members from unrelated Galactic field stars or extragalactic contaminants. The resulting samples of 'MYStIX Probable Complex Members' (MPCM) represent the largest census yet obtained for most of the target star forming regions.

The MYStIX approach has its limitations; some stellar subpopulations are poorly recovered, and it is difficult to establish the completeness limit of the combined samples. MPCM samples have substantial limitations (Appendix B) but give more comprehensive samples than any single-method approaches to uncovering young stellar populations (optical variability, Hα surveys, K-band excesses, mid-infrared excesses, X-ray emission, spatial clustering). For MSFRs with high levels of nebular emission and/or Galactic field star contamination, often no serious attempts had been made to define individual complex members.

The MPCM samples thus represent the largest and most comprehensive membership lists for most MYStIX regions. These samples are particularly advantageous by their inclusion of pre-main sequence stars both with and without protoplanetary disks. Disk-free systems are found from X-ray surveys that efficiently remove contamination by older Galactic field stars; this considerably extends our view of past star formation (stellar ages > 2 Myr) in these regions. Furthermore, our fields of view are relatively large, typically 5 to 30 pc in extent, permitting a view of star formation in the vicinity of rich clusters. Finally, the MYStIX study has the advantage of a consistent treatment of a significant number of young stellar clusters. This permits comparisons to examine how various cluster properties (such as spatial extent and shape, monolithic vs. clumpy structure, mass segregation, triggered star formation, spatial-age gradients, and OB wind interactions with clouds) appear under differing conditions.

The astrophysical issues outlined in §1.1 can be powerfully addressed using MPCM samples. In the two prototype fields examined here, NGC 2264 and Trifid Nebula, younger disk-bearing and older disk-free stars show different spatial structures suggesting that star formation in these regions has a complicated history likely involving multiple star formation sites over an extended period of time. Future MYStIX science studies will include: a multifaceted spatial study of clustering to elucidate dynamical states; a new stellar age estimator applied to (sub)clusters to elucidate star formation histories; a search for previously unremarked massive members; study of triggered star formation in molecular cloud material adjacent to rich clusters; investigation of OB winds both close to and far from the stellar surface; comparison of cluster IMFs; and other issues. These issues will be discussed in a series of forthcoming papers.

Acknowledgements: We thank J. Forbrich and P. Teixeira (Univ. Vienna) for useful discussion about NGC 2264. The MYStIX project is supported at Penn State by NASA grant NNX09AC74G, NSF grant AST-0908038, and the Chandra ACIS Team contract SV4-74018 (G. Garmire & L. Townsley, Principal Investigators), issued by the Chandra X-ray Center, which is operated by the Smithsonian Astrophysical Observatory for and on behalf of NASA under contract NAS8-03060. M. S. Povich was supported by an NSF Astronomy and Astrophysics Postdoctoral Fellowship under award AST-0901646. This research made use of data products from the Chandra Data Archive and the Spitzer Space Telescope, which is operated by the Jet Propulsion Laboratory (California Institute of Technology) under a contract with NASA. The United Kingdom Infrared Telescope is operated by the Joint Astronomy Centre on behalf of the Science and Technology Facilities Council of the U.K. This work is based in part on data obtained as part of the UKIRT Infrared Deep Sky Survey and in part on data obtained in UKIRT director's discretionary time. This research used data products from the Two Micron All Sky Survey, which is a joint project of the University of Massachusetts and the Infrared Processing and Analysis Center/California Institute of Technology, funded by the National Aeronautics and Space Administration and the National Science Foundation. The HAWK-I near-infrared observations were collected with the High Acuity Wide-field K-band Imager instrument on the ESO 8-meter Very Large

Facilities: CXO (ACIS), Spitzer (IRAC), CTIO:2MASS (), UKIRT (WFCAM)

## A. MYStIX Star-Forming Complexes

This Appendix gives brief overviews of the MYStIX MSFRs based on previous studies of their stellar populations. The regions are listed in order of increasing distance (Table 1). Comprehensive descriptions of the regions based on pre-2008 observations are provided in the Handbook of Star Forming Regions (Reipurth & Schneider 2008).

### A.1. Orion Nebula

The stellar content of the Orion Nebula (M 42) is the best studied for any massive star formation region in the sky (Muench et al. 2008). The region of interest for MYStIX is a single Chandra ACIS $17' \times 17'$ field centered on the Orion Nebula Cluster (ONC). The ONC, or the Orion Id OB association, is a monolithic, centrally condensed rich cluster of stars with about 3000 members down to the stellar limit. The stellar distribution can be modeled as an isothermal ellipsoid elongated north-south with core radius 0.2 pc and central star density $2 \times 10^4$ stars $pc^{-3}$. The cluster exhibits strong mass segregation with the 'Trapezium' of OB stars concentrated inside the core dominated by the O7 star $\theta^1$C Ori with mass around 30 $M_\odot$. The typical age of ONC stars is around 2 Myr with controversial evidence for a wide age spread over $1 - 10$ Myr. The ONC is superposed on a $1°$-long molecular filament along the center of the Orion A cloud with current star formation in its cores. The OMC-1 core lie in the MYStIX field with two components: the Becklin-Neugebauer Kleinman-Low region and the OMC-1S core. They contain several dozen embedded protostars, including several with high mass, seen in infrared and X-ray wavelengths.

The ONC stellar population has been subject to three recent intensive surveys with major space telescopes. First, the Chandra Orion Ultradeep Project (COUP) observed the Nebula for 13.2 nearly-contiguous days, producing a catalog of 1400 X-ray emitting young stars (Getman et al. 2005). Second, an intensive survey with several instruments on the Hubble Space Telescope and associated ground-based telescopes produced a sensitive high-resolution catalog of cluster members (Da Rio et al. 2009; Robberto et al. 2010). Third, a multi-epoch Spitzer mid-infrared survey has found periodic or aperiodic variability in 1200 ONC stars (Morales-Calderón et al. 2011).

### A.2. Flame Nebula

The Flame Nebula, or NGC 2024, is a prominent H$\textsc{ii}$ region in the L 1630 (Orion B) dark cloud near the Orion Belt star $\zeta$ Ori and the Horsehead Nebula (Meyer et al. 2008).

Although it is the richest star cluster in the Orion complex after the ONC, visible band studies are impeded by a dark lane of cloud material obscuring the cluster. A ridge of dense molecular cores lie behind the cluster. A thorough infrared census of the cluster has not yet been reported and the spectral types of the dominant stars are not well-established. The fraction of members with infrared excess protoplanetary disks appears to be high around $70 - 80\%$. Chandra images show $250$ cluster members with mean absorption around $A_V$ 10 mag.

## A.3. W 40

The W 40 H$_{II}$ region and associated cluster, though close to the Sun, was poorly studied until recently due to high local obscuration (Rodney & Reipurth 2008). At visible wavelengths, the ionized nebula is seen but the central cluster is covered with a screen of dusty cloud, similar to the Flame Nebula. Several late-O or early-B stars powering the region were identified at radio and infrared wavelengths. The Chandra source catalog gives $200$ cluster members with $A_V$ $5 - 20$ mag obscuration and an estimated $600$ star total population (Kuhn et al. 2010). The near-infrared disk fraction around 50% implying an age $< 1$ Myr. The structure appears roughly spherical with core radius $0.15$ pc. Mass segregation is seen; not only are massive stars concentrated into the core, but stars below 1.5 $M_\odot$ more dispersed than intermediate-mass stars. Star formation is not present in a small molecular core just west of the cluster, but mid-infrared protostars detected with Herschel are prevalent in the obscuring dust lane and elsewhere in the vicinity (Maury et al. 2011).

## A.4. RCW 36

RCW 36 is the smallest, and presumably the youngest, of several H$_{II}$ regions distributed over $10°$ in the Galactic Plane in the Vela Molecular Ridge Cloud C (Pettersson 2008). Infrared imaging shows a cluster with $> 350$ members within a radius of 0.5 pc; the central density is $3000$ stars pc$^{-2}$. The two brightest stars have spectral types O9 and the distance is estimated to be 700 pc (Ellerbroek et al. 2012). The cluster is obscured with typical $A_V$ 10 mag. The structure shows two clumps separated by $0.2$ pc; massive stars appear concentrated in the northern clump. The cluster illuminates several bright rimmed clouds in the nearby cloud, and a large area ($3$ pc in extent) of heated dust. Herschel images reveal dense filament of cold dust with $A_V > 100$ mag lies within the nearby cloud, suggesting that more massive star formation may occur in the future (Hill et al. 2011).

## A.5. NGC 2264

NGC 2264 (Christmas Tree Cluster) associated with the Cone Nebula bright rimmed cloud and Fox Fur Nebula H$_{II}$ region has low contamination by Galactic field stars and negligible interstellar absorption to the region. Known as an emission nebula since the 18$^{th}$ century, it was the site where Merle Walker first described stars during their pre-main sequence phase during the 1950s (Dahm 2008a). The most massive stellar member is the O7+O9.5 main sequence binary S Mon; and at least 30 B stars are present. It does not have a monolithic cluster structure, but rather appears to be a collection of $2-3$ clusters with additional distributed young stars. A true age spread is probably present among the lower mass stars ranging from protostars around IRS 1 and IRS 2 to several hundred X-ray selected stars, many of which are disk-free (Class III). A considerable number of Class II and Class III stars are dispersed $5-10$ pc from the currently active star forming clouds. A new infrared study estimates that the total stellar population is 1400 stars (Teixeira et al. 2012). A deep 300 ks Chandra observation of the southern portion of the region is now underway.

## A.6. Rosette

The Rosette Nebula is ionized by NGC 2244, the youngest cluster within the large Mon OB2 association. It produces a blister H$_{II}$ region on an edge of the Rosette Molecular Cloud that extends east of the nebula (Román-Zúñiga & Lada 2008). The cloud consists of several clumps spread over 1.5°, imaged with the Herschel satellite (Schneider et al. 2010). A linear mosaic of five Chandra fields cover the cluster and, with lower sensitivity, portions of the cloud (Wang et al. 2010, and references therein). The central cluster is quite rich with over thirty stars earlier than B3, dominated by two 100 M$_\odot$ stars. The OB stars do not show mass segregation with respect to the pre-main sequence stars. Two clusters on the periphery of NGC 2244, RMS XA and NGC 2237 each with 200 stellar members, may be have been triggered by the main cluster in cloud material that is now dissipated. Several smaller clusters are embedded in the cloud, well-populated in both X-ray and infrared surveys. Although the cloud dust is heated by the central cluster massive stars, most of the embedded clusters do not appear to be triggered by the expanding H$_{II}$ region.

## A.7.  Lagoon Nebula

The Lagoon Nebula (M 8) is a large ($\sim 50' \times 40'$) H$_{II}$ region in the nearby Sagittarius-Carina spiral arm in front of the Galactic Center region (Tothill et al. 2008). NGC 6530, the central cluster, is dominated by the O4 star 9 Sgr with nearly 70 other OB stars. A secondary nebular region called the Hourglass Nebula is powered by the O7 star Her 36. The H$_{II}$ region contains numerous bright rimmed clouds, pillars, and molecular clumps. Due to heavily contamination by Galactic field stars, the cluster can barely be discerned in the distribution of optical or 2MASS stars. A Chandra study finds $\sim 800$ members, centrally concentrated with core radius around 1.6 pc and extending beyond 7 pc. Mass segregation is present. The pre-main sequence stars have typical ages between $\sim 0.8$ and 2.5 Myr, and near-infrared disk fraction appears to be high. In the surrounding molecular cloud, over 60 Class 0/I protostars are identified by mid-infrared excess in Spitzer images, indicating active current star formation (Kumar & Anandarao 2010). Her 36 is surrounded by a distinct concentrated Hourglass Nebula Cluster with at least 100 stars, many of them with near-infrared disks.

## A.8.  NGC 2362

Along with Tr 15 in Carina, NGC 2362 may be the oldest cluster in the MYStIX sample with age $\sim 5$ Myr. It is dominated by the O9 Ib supergiant $\tau$ CMa. No main sequence O stars are present, likely lost by supernovae, and $\sim 40$ B stars are present. Molecular material is absent immediately around the cluster, although triggered star formation in more distant clouds may be active. With no obscuration and a location several degrees off the Galactic Plane, field star contamination is not heavy and membership of $\sim 300$ stars can be established in optical color-magnitude diagrams. About a third of these are H$\alpha$ emitters. Nearly 400 X-ray sources are seen in the Chandra field; the total stellar population is below that of the ONC. Infrared excesses from full protoplanetary disks are rare, although depleted and transition disks are more common (Currie et al. 2009).

## A.9.  DR 21

The DR 21 star forming region is part of the huge Cygnus Super-Bubble (Cygnus X, $\sim 15°$ in size; Reipurth & Schneider 2008). It lies at the end of an unusually dense molecular filament several parsecs in length with embedded high-mass young stellar objects producing masers. The DR 21 cloud itself has a deeply embedded massive star producing an ultra-compact H$_{II}$ region and molecular outflow; however, it is not accompanied by a rich cluster.

It is possible that the cluster has not yet formed; the cloud has    20,000 $M_0$ of molecular gas within    1 pc, but may be supported against collapse by magnetic and turbulent pressure (Kirby 2009). Several dozen mid-infrared excess young stars in Spitzer images are distributed in the cloud and along the molecular filament (Kumar et al. 2007). The Chandra X-ray findings of the region have not been published prior to MYStIX.

## A.10.   RCW 38

RCW 38 is an unusually young, heavily obscured rich cluster that appears associated with, but probably lies behind, the Vela Molecular Ridge (see RCW 36 above; Wolk et al. 2008). The cluster is dominated by a heavily obscured O4 star accompanied by an estimated    30 additional OB stars. Some are strongly concentrated in the cluster core while others are dispersed. The identified pre-main sequence population has over    600 disk-bearing stars with infrared excesses, many of which are seen in a Chandra X-ray image (Winston et al. 2011). Obscuration varies widely from $A_V$    3 to 60 magnitudes. The X-ray sources show several subclusters spread over several parsecs in addition to a central dense concentration.

## A.11.   NGC 6334

The NGC 6334 complex, or the Cat's Paw Nebula, is a major star forming region on the Sagittarius-Carina spiral arm close to the NGC 6357 complex (Persi & Tapia 2008). Several luminous mid-infrared sources in the central $10^|$ aligned along the Galactic Plane mark very young embedded clusters. NGC 6334 I(N) is a proto-massive star well-studied at millimeter and far-infrared wavelengths. Optical and infrared images are dominated by filamentary ionized gas and heated dust, dense patchy obscuration, and heavy contamination by Galactic field stars. Hence there is no catalog of cluster members, or even OB stars, in the region. A mosaic of Chandra fields reveals several unobscured star clusters in front of and around the dense cloud, as well as members of the embedded clusters (Feigelson et al. 2009). The X-ray source population is rich with > 1500 cluster members.

## A.12.   NGC 6357

NGC 6357 appears to have formed from the same giant molecular cloud as NGC 6334 (Russeil et al. 2010). G353.2+0.9 is its brightest H$_{II}$ region on the northern rim of an annular evacuated region in the cloud, ionized by the massive stellar cluster Pismis 24 (Bohigas et al. 2004). The three most massive (O3) stars in Pis 24 each have masses 100 $M_\odot$ (Maíz Apellániz et al. 2007). NGC 6357's degree-sized shell seen in H$\alpha$ may outline a superbubble blown by a MSFR that preceded Pis 24. The presence of a post-main sequence Wolf-Rayet star inside this shell, and the possible X-ray discovery of an older population dispersed around Pis 24 (Wang et al. 2007), provide indirect evidence that the cavity supernova remnants of an older cluster might have helped to expand the bubble into the 60' shell structure seen today. Chandra observations reveal previously unrecognized embedded clusters in the molecular cloud south and east of Pis-24. The full region may be one of the closest and youngest examples of a giant molecular cloud complex engaged in rapid, extensive, nearly coeval, multiple massive stellar cluster formation (Townsley et al. in prep.). This makes NGC 6357 a rare addition to the Galaxy's complement of "clusters of clusters" – a mode of star formation that appears to be inherently different than the more familiar single, monolithic cluster formation that created such regions as M 17 or NGC 3603. NGC 6357 may represent an earlier phase of older "cluster of cluster" complexes like the Carina Nebula and NGC 604 in M 33.

## A.13. Eagle Nebula

The Eagle Nebula (M 16) and its ionizing cluster NGC 6611 have been popular targets for multiwavelength star formation studies since the famous Hubble Space Telescope images of its "Pillars of Creation" (Oliveira 2008). It contains 13 O stars, at least half of which are binary (Sana et al. 2009), and may have lost several more through dynamical ejections (Gvaramadze & Bomans 2008). Its earliest stars are an O3.5 V and an O4 V, both in binary systems. A Chandra ACIS-I pointing towards NGC 6611 revealed 1,101 X-ray point sources (Linsky et al. 2007); a recent re-analysis of this dataset (Flagey et al. 2011) also shows faint diffuse X-ray emission that these authors suggest might be produced by a cavity supernova. Adding two more ACIS-I pointings on the eastern side of the complex to the original NGC 6611 ACIS data, Guarcello et al. (2012) catalog 1,755 X-ray point sources.

## A.14. M 17

M 17 produces the second-brightest H II region in the sky (Chini & Hoffmeister 2008), and its massive central cluster NGC 6618 is very young with >8000 total members (Broos et al. 2007). The central O4+O4 binary shows evidence in the X-ray (Broos et al. 2007), IR (Hoffmeister et al. 2008), and radio (Rodríguez et al. 2012) for being a pair of colliding-wind binaries, implying that the region is ionized by at least 4 early-O stars. It is situated at the edge of one of the Galaxy's most massive and dense molecular cloud cores, M 17 SW, at a distance of 2.0 kpc (Xu et al. 2011). Its orientation gives an excellent view of the interface between the HII region and the molecular cloud, and of the outflow of shocked massive stellar winds into the Galactic interstellar medium (Meixner et al. 1992; Townsley et al. 2003; Pellegrini et al. 2007). Near this young massive cluster, a large bubble to its north hosts a 2–5 Myr old young stellar population that may represent the previous generation of star formation in the M 17 complex (Povich et al. 2009). NGC 6618 is one of the few bright massive star-forming regions that has sufficient stellar wind power to produce a bright X-ray outflow (Townsley et al. 2011b) and yet is unlikely to have hosted any supernovae.

## A.15. W 3

W 3, at the western side of the W3-W4-W5 complex, is by itself an important complex in the outer Galaxy. It exhibits the full range of massive star formation environments fueled by material from the "high density layer" where the W4 superbubble (see below) is inter-acting with its adjacent giant molecular cloud (Megeath et al. 2008). The W4/W3/HB3 complex contains one of the most massive molecular clouds in the outer Galaxy (Heyer & Terebey 1998), massive embedded protostars (Turner & Welch 1984; Megeath et al. 1996), every known type of radio HII region (Tieftrunk et al. 1997), and one of the largest super-nova remnants in the Galaxy (Routledge et al. 1991). A prominent monolithic, revealed, centrally-concentrated cluster just east of the W 3 cloud is the 3–5 Myr old IC 1795 (Roc-catagliata et al. 2011). W3 North is a parsec-scale HII region powered by the isolated O7 star IRAS 02230+6202; Chandra observations establish that it has no lower-mass accompanying population (Feigelson & Townsley 2008). W3 Main is a rich, centrally concentrated embed-ded cluster characterized by sequential star formation (Bik et al. 2012). It is still forming massive stars, revealed by hypercompact radio HII regions, but it also hosts a rich, older pre-main sequence population that has lost most of its protoplanetary disks (Feigelson & Townsley 2008). Further to the south, W 3(OH) is a well-studied ultra-compact HII region ionized by an O9 star (Hirsch et al. 2012) and surrounded by a cluster of more than 200 stars (Carpenter et al. 2000). OH and $H_2O$ masers, molecular outflows, and strong CO line emission indicate the presence of massive embedded protostars.

## A.16. W 4

The radio continuum and Hα nebula W 4 is a 14 Myr old (Lagrois & Joncas 2009) superbubble in the Perseus Arm of the Milky Way, perhaps the nearest interstellar 'chimney' between the dense Galactic plane gas and the Galactic halo. Although too young to have formed the W4 superbubble (Lagrois et al. 2012), IC 1805 is a 1–3 Myr old (Massey et al. 1995) massive young stellar cluster now re-energizing the preexisting gaseous structure. The single Chandra pointing of IC 1805 is centered on HD 15558, a massive binary or possibly a triple system (De Becker et al. 2006). This cluster contains a large number of intermediate-mass stars that exhibit a wide range of disk properties (Wolff et al. 2011).

## A.17.  Carina Nebula

A Chandra mosaic of 22 ACIS-I pointings of 60 ks each revealed over 14,000 X-ray point sources and extensive diffuse emission (Broos et al. 2011a; Townsley et al. 2011b). These complement extensive near-infrared observations with the VLT's HAWK-I instrument (Preibisch et al. 2011) and mid-infrared mapping with Spitzer's IRAC instrument (Smith et al. 2010; Povich et al. 2011a). Many of these studies are collected into the Chandra Carina Complex Project (Townsley et al. 2011a).

The MYStIX project adopts the portion of the Chandra mosaic that is covered by both HAWK-I and Spitzer. This field includes Tr 14, Tr 15, and Tr 16 that are the most massive clusters in a system of many clusters and stellar groups that form the Carina star-forming complex (Feigelson et al. 2011). While Tr 14 and Tr 15 are centrally-concentrated and mass-segregated, Tr 16 consists of several clumps (Wolk et al. 2011) and appears some-what older than Tr 14 with evolved supergiants including the remarkable Luminous Blue Variable, η Car. Tr 15 appears yet older and appears to have lost its most massive stars as supernovae (Wang et al. 2011). An isolated neutron star discovered in XMM observations of Carina (Hamaguchi et al. 2009; Pires et al. 2012) and bright complex diffuse X-ray emission (Townsley et al. 2011b) strengthen the case for extensive supernova activity in the Carina complex. The Herschel satellite observations reveal large quantities of atomic and molecular material remain in Carina, particularly in the 'South Pillars' region, with densities sufficient to continue fueling star formation (Preibisch et al. 2012).

## A.18.  Trifid

The Trifid Nebula (M 20) is an optically bright H II region trisected by three dust lanes (Rho et al. 2008). The main emission nebula is ionized by the main sequence O7 star HD 164492A that lies in a dense group of intermediate-mass stars. An evolved supergiant heats a reflection nebula to the north. A molecular cloud surrounds most of the H II re-gion, fragmented into dense cores with $> 30$ embedded Class 0/I protostars that are widely distributed along the dust lanes and in the surrounding molecular cloud. Bright rimmed clouds are associated with star formation. The stellar population includes $> 80$ candidates identified by $K$-band photometric excess, $> 150$ with mid-infrared excess, and 300 X-ray sources. The distance is uncertain and estimates have recently increased from 1.7 kpc to 2.7 kpc; it thus is probably not in the Sagittarius-Carina arm like the Lagoon and Eagle Nebulae.

## A.19. NGC 3576

This massive star-forming complex contains at least two major clusters, both prominent in X-rays (Townsley et al. 2011b): a very young, massive, embedded cluster ionizing a giant H$\scriptstyle\rm II$ region (de Pree et al. 1999), and an older, revealed, more relaxed cluster to its north. The southern cluster is so deeply embedded that its dust-processed infrared emission saturates most detectors. It contains $>50$ OB stars (Maercker et al. 2006) but the census of its ionizing sources is still incomplete (Figuerêdo et al. 2002; Barbosa et al. 2003). The embedded cluster shows a plume of hot X-ray-emitting plasma just breaking through the edge of its giant molecular cloud; this outflow may be similar to that seen in M 17, but with a less convenient face-on orientation and earlier in its evolution (Townsley et al. 2011b). The northern revealed cluster is not well-studied; its two most massive members (late-O stars) constitute the NGC 3576 OB Association (Humphreys 1978), but the accompanying young cluster (ASCC 65) was only recently recognized (Kharchenko et al. 2005). A young pulsar, PSR J1112-6103 (Manchester et al. 2001), is situated near the center of ASCC 65; if it lies at the same distance, it may be the remains of one of this cluster's more massive members (Townsley et al. 2011b). ASCC 65 may resemble Tr 15 in Carina: a massive cluster with age $5-10$ Myr where the IMF is truncated at high mass by the evolution and supernova of its most massive stars.

## A.20. NGC 1893

The young star cluster NGC 1893 and is associated H$\scriptstyle\rm II$ region IC 410, a portion of the Aur OB2 association, have only recently been studied. Lying towards that Galactic anti-center at a distance 3.6 kpc, it is in the MYStIX sample due to a very deep Chandra observation as well as Spitzer, near-infrared, optical, and H$\alpha$ imaging observations (Caramazza et al. 2008; Prisinzano et al. 2011). They reveal 360 cluster members; the majority are Class II systems with infrared excesses, although a few are protostars. Most members have estimated ages below 2 Myr. The cluster has evacuated a large interstellar region and little molecular material lies within the Chandra field of view.

## B.   Limitations of the MPCM sample

Section 5 and the accompanying papers (Kuhn et al. 2013a; Townsley et al. 2013; King et al. 2013; Kuhn et al. 2013b; Naylor et al. 2013; Povich et al. 2013; Broos et al. 2013) describe the construction of the MYStIX Probable Cluster Member (MPCM) samples for the MYStIX MSFRs listed in Table 1. The MPCM samples are based on analysis, both individually and combined, of Chandra X-ray, UKIRT and 2MASS near-infrared, and Spitzer mid-infrared imaging observations. Based on our results and validation procedures (§6-7), we discuss here a variety of uncertainties and limitations of the MPCM samples.

Spurious sources in MYStIX X-ray source samples By pushing down to 3 photon sources and sub-arcsecond resolution of double sources on-axis (Kuhn et al. 2013a; Townsley et al. 2013), we increase the possibilities that these faint and proximate sources do not exist. However, this problem is likely not severe. In the Chandra Carina Complex Project where the X-ray detection procedure is the same as with MYStIX and superb near-infrared imaging is available from the HAWK-I camera on ESO's Very Large Telescope, 89% of the X-ray sources have counterparts, 93% of which are classified as probable members of the star forming region using a statistical classifier (Broos et al. 2011a; Preibisch et al. 2011). The MYStIX X-ray and classification methods are closely modeled on those of the CCCP. We believe that only a few percent of the X-ray sources are likely to be spurious, and these are unlikely to be matched to infrared sources and be successfully classified as members by the statistical classifier.

Spurious sources in MYStIX near-infrared and mid-infrared source catalogs We have tuned the analysis to optimize the detection and photometry of faint stellar sources in the presence of moderate levels of crowding and nebulosity (King et al. 2013; Kuhn et al. 2013b). Thus, the catalogs have few missing detections (false negatives) when compared to visual examination of lightly contaminated images. But the analysis is not optimized for the elimination of spurious sources. The false positives have several origins; reducing their number using an automated algorithm without also reducing sensitivity is difficult. However, the false positives rarely enter the MPCM samples, as they are unlikely to be positionally matched with X-ray sources or satisfy the stringent SED criteria for infrared-excess. We are more concerned, in cases where the Galactic field star contamination is high, about the chances that some real infrared excess sources are red giant stars with dusty envelopes are classified as MSFR members; conservative criteria involving downweighting of widely dispersed infrared-excess stars are used to reduce this source of contamination (Povich et al. 2013).

**Incompleteness in MYStIX mid-infrared source catalogs** When mid-infrared nebular emission (primarily from PAH molecules) is strong and spatially variable, the sensitivity of any algorithm seeking unresolved stellar sources is reduced (Kuhn et al. 2013b). In addition, as the Spitzer telescope has several-fold lower resolution than the (on-axis) Chandra or UKIRT telescopes, source confusion in the central regions of rich clusters can produce incorrect photometry and missing sources. This is not a small effect: central regions of Tr 14 in Carina, NGC 3576, NGC 6618 in M 17, W 3 Main in W 3, and other clusters are seriously deficient in mid-infrared sources for these reasons. This systematic deficiency in mid-infrared sources will cause spatial biases in the census of disk-bearing young stars.

**Incorrect counterparts in MYStIX multiwavelength matching** The matching problem inherently has no ideal solution for difficult cases such as off-axis Chandra sources (with large positional errors due to telescope coma) associated with crowded Galactic Plane infrared fields. Some statistical technique, such as our probability scaled to K band magnitudes (Naylor et al. 2013), is needed and will necessarily give some incorrect counterparts. This matching problem has negligible scientific impact for nearby uncrowded fields like NGC 2264 (providing one treats unresolved binary systems as single stars). But it is a potentially serious problem for MYStIX targets like the Trifid, Lagoon, NGC 6357 and NGC 6334 with Galactic longitudes $|l| \leq 10°$.

**Unreliability of MYStIX X-ray source classifications** The probabilistic nature of our assignment of X-ray sources to the MPCM samples (Broos et al. 2013) will unavoidably produce false negatives and false positives. The class likelihood distributions for J, X-ray median energy, and X-ray spatial distributions may be inaccurate or ineffective for the discrimination of young stars from contaminant populations. The decision rule for assignment into classes is to some degree arbitrary, and reasonable differences in classification procedure will give different MSFR membership lists. This problem is likely to be quantitatively unimportant for nearby and lightly contaminated MSFRs like NGC 2264, but becomes important for distant, crowded clusters like the Trifid. Until spectroscopic followup studies are conducted for MPCM samples, we cannot quantitatively evaluate the importance of this problem. The high success rate of MSFR membership lists derived from Chandra/HAWK-I and Chandra/Spitzer matching in the CCCP (Povich et al. 2011b; Preibisch et al. 2011) in the Carina Nebula suggests that this problem is not severe. We provide electronic tables enabling other scientists to apply different choices of classification criteria.

**Bias against widely distributed young stars** At two steps in the MYStIX analysis process – in the weighting of young stellar disks to the local surface density of infrared-excess stars (§5.5), and in the weighting of MPCM classification by the local surface density of X-ray sources (§5.7) – our procedures generating the MPCM lists favor spatially concentrated members and disfavor widely dispersed young stellar populations. The MPCM samples thus can not give reliable quantitative insight into the ratio of clustered vs. distributed star formation.

**Bias against intermediate-mass stars** The census of young intermediate-mass A and late-B stars may be systematically incomplete in MPCM samples. First, infrared-excess criteria may miss a larger fraction of young AB stars than lower-mass stars because protoplanetary disks evolve faster at higher stellar masses (Carpenter et al. 2006). Second, X-ray samples miss many young AB stars because intrinsic X-ray emission is weak or absent, although some are found due to lower mass companions (e.g., Berghöfer et al. 1997). MPCM samples may thus exhibit a spurious drop in the IMF at intermediate masses.

**Bias against very deeply embedded stars** The census of the most deeply embedded stars, say $50 \leq A_V \leq 200$, in the MPCM samples is deficient both due to loss of X-ray photons from soft X-ray absorption, and to erroneous classification as extragalactic or 'unclassified' sources. These sources would have Chandra X-ray median energies in the range $3 < MedE < 6$ keV (the calibration of median energy and interstellar absorption is discussed by Getman et al. 2010). The misclassification arises from an operational limitation of our statistical classifier because the training sets of young stars do not sufficiently populate the high-absorption tail of the median energy distribution. Some of these heavily obscured stars will be protostars captured in the MIRES sample as IR-excess stars, but the comparison with Class I protostars in NGC 2264 Forbrich et al. (2010) shows that many will be missing. Others will be missed due to nebular contamination and crowding in the infrared images. A detailed study of MYStIX protostars would thus benefit by recovering X-ray sources with high median energies and infrared excess stars that were not classified as MPCMs.

**Difficulties in establishing completeness limits** As MYStIX combines X-ray, near-infrared and mid-infrared data in a complicated fashion, no straightforward statement of sensitivity limits can be presented. X-ray sensitivity limits are strongly affected by off-axis degradation of Chandra mirror performance, and by variable exposure times in overlapping exposures of MYStIX mosaics. This can be mitigated by defining a subsample of 'spatially complete' X-ray sources; in the CCCP, this truncation removed about 2/3 of the X-ray sources from consideration (Feigelson et al. 2011). Even for this spatially uniform subsample, luminosity and mass limits depends on line-of-sight absorption. Infrared-excess sensitivity limits are roughly constant in regions without nebulosity, but are badly degraded within PAH-bright HII regions. Conceptually, the infrared excess selection criterion discovers protoplanetary disks not stars, and a well-defined completeness limit for disks does not give a clear completeness limit for the host stars. The published samples of OB stars confirmed by optical spectroscopy have no clear completeness limits, and probably differ strongly among the MYStIX targets. As MPCM samples combine these X-ray, infrared-excess, and OB datasets, the completeness of the resulting sample can not be evaluated.

**MYStIX samples do not contain all known young stars** The stellar populations of a few MYStIX clusters have been extensively studied in Hα and stellar variability surveys. The Orion Nebula Cluster has been very carefully surveyed, and high-quality Hα surveys are available for NGC 2264, NGC 2362, DR 21, and the Flame Nebula. These historical samples have not been incorporated into the MYStIX analysis because they are not uniformly available for all targets. Many Hα and variable stars are independently captured by the MYStIX survey procedures (see §7 above and footnotes in electronic tables of Broos et al. 2013), but others are not recovered. We also have not included far-infrared, submillimeter and millimeter surveys of Class 0-I protostars obtained with telescopes such as the Herschel satellite, James Clark Maxwell Telescope, Submillimeter Array, or Atacama Large (Sub)Millimeter Array. Science analysis can thus often be enhanced by combining MPCM with other published samples of young stars.